\documentclass[times,twocolumn,final]{elsarticle}

\usepackage{medima}

\usepackage{amssymb}
\usepackage{amsmath}
\usepackage{cases}
\usepackage{bbm}
\usepackage{latexsym}

\usepackage{url}
\usepackage{xcolor}

\usepackage{hyperref}

\definecolor{newcolor}{rgb}{.8,.349,.1}

\journal{ }

\begin{document}

\verso{Ke Lei \textit{et~al.}}

\begin{frontmatter}

\title{Artifact- and content-specific quality assessment for MRI with image rulers}

\author[1]{Ke \snm{Lei}\corref{cor1}}
\ead{kllei@stanford.edu}
\cortext[cor1]{Corresponding author: }
\author[1]{John M. \snm{Pauly}}
\author[2]{Shreyas S. \snm{Vasanawala}}

\address[1]{Department of Electrical Engineering, Stanford University, Stanford, CA, USA}
\address[2]{Department of Radiology, Stanford University, Stanford, CA, USA}

\received{June 2021}

\begin{abstract}
In clinical practice MR images are often first seen by radiologists long after the scan. If image quality is inadequate either patients have to return for an additional scan,  or a suboptimal interpretation is rendered. An automatic image quality assessment (IQA) would enable real-time remediation. 
Existing IQA works for MRI give only a general quality score, agnostic to the cause of and solution to low-quality scans. Furthermore, radiologists' image quality requirements vary with the scan type and diagnostic task. Therefore, the same score may have different implications for different scans. 
We propose a framework with multi-task CNN model trained with calibrated labels and inferenced with image rulers. Labels calibrated by human inputs follow a well-defined and efficient labeling task. Image rulers address varying quality standards and provide a concrete way of interpreting raw scores from the CNN. 
The model supports assessments of two of the most common artifacts in MRI: noise and motion. It achieves accuracies of around 90\%, 6\% better than the best previous method examined, and 3\% better than human experts on noise assessment. Our experiments show that label calibration, image rulers, and multi-task training improve the model's performance and generalizability. 

\end{abstract}

\begin{keyword}
\KWD Image quality assessment\sep Multi-task learning\sep Artifact detection\sep Convolution neural networks\sep MRI
\end{keyword}

\end{frontmatter}

\section{Introduction}
Magnetic resonance imaging (MRI) can provide detailed images with clear contrast, but unfortunately with long scan times. Various categories of image artifacts degrade diagnostic image quality, with varied causes and remedies. Some of the common artifacts are heavily patient dependent. Therefore, no preset scan parameters can guarantee satisfactory image quality. When suboptimal images are reviewed for interpretation after the image acquisition has been completed, either the patient is asked to return for a repeat scan, or a limited interpretation is rendered. Ideally, if all images are checked before the patient leaves the scanner, repeat scans and limited interpretations can be minimized. However, in busy clinical settings, MR technologists with varied skill levels do not reliably assure image quality. Even if the technologists do have the time to check the images, these operators may not know how high the image quality must be for accurate interpretation nor how to improve the image quality. This poses two questions.  First, what is adequate quality? Second, what is the specific cause of inadequate quality? 

The answer to the first question varies with scan types and diagnostic tasks. Fat-suppressed (FS) scans normally have a lower signal-to-noise ratio (SNR) than their counterpart, non fat-suppressed (NFS) scans. Because the strong but irrelevant signals from fat regions are suppressed, this image type is often critical for diagnosis,  so radiologists can accept a higher noise level than in non-fat suppressed scans. If the diagnosis relies on a small or subtle abnormality higher quality is needed than when the relevant feature relies is readily apparent. 
Image quality requirements also depends on the body part being examined. For example, spatial resolution requirements in imaging a knee for a sports injury may be higher than an abdominal evaluations. Higher levels of motion artifacts are tolerated in abdominal imaging compared with say brain imaging.
We aim to have an automatic framework that assists operators with ensuring image quality and ultimately takes this responsibility without human attention. Therefore, we need a systematic way of answering the first question. We propose content-specific (i.e. scan-specific) image rulers to capture the quality preferences from radiologists and make that information systematically interpretable by other human and models.   

\begin{figure}[!t]
\label{artifacts}
\centering
\includegraphics[width=0.97\linewidth]{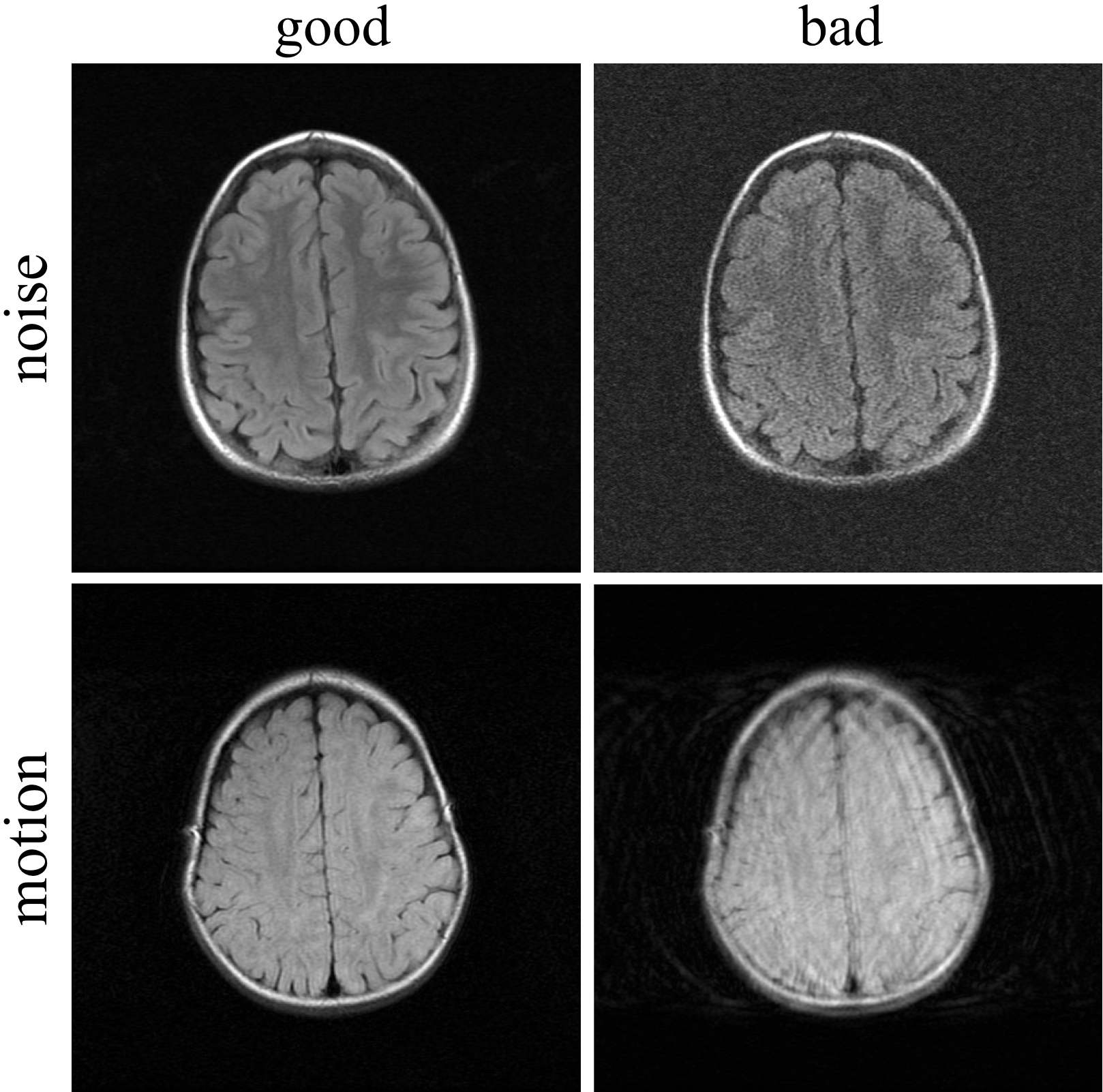}
\caption{Samples of noise and motion artifacts in MRI.}
\end{figure}

We need the answer to the second question so that operators (MRI technologists) can take specific and effective action to improve the image quality. Noise and motion are two types of artifacts (Fig. \ref{artifacts}) that exist in all in-vivo MR scans. The noise content depends on factors such as body shape and characteristics, scan time, coil geometry, etc. The most controllable factor is scan time. Theoretically, the SNR of the measurement increases linearly with square root of scan time. Therefore, if a scan is determined to be too noisy, a longer scan with a larger number of signal averages can be prescribed for a repeat scan. Ultimately, instead of assessing the perceptual noise level at the end of a scan, models running in real-time could monitor a scan as it is in process and stop it as soon as desired quality is obtained. In addition to avoiding a repeat scan, excess scan time is eliminated. 
The presence of motion depends on many factors, including the body part, the patient's ability to hold still or suspend respiration, and the motion-robustness of the type of scan. If a scan is determined to have too much motion artifact, the operator can coach the patient or choose a more motion-robust sequence. 
The framework presented here checks for both types of artifact in parallel. From the outcome of each check, the operator can know if a scan is satisfactory, and if not, the reason that it failed.  

\vspace{1.7mm}
\noindent\textbf{Related work.} There are prior reports on estimating the degree of noise in natural images, MRI, and computed tomography (CT) using principal component analysis on local patches \citep{Liu,MANJON201535,Chen}, estimators on specific noise distribution models \citep{AJAFERNANDEZ2015184}, and statistical evaluation of discrete cosine transforms (DCT) of local patches \citep{iedd} or local signal variances of filtered images \citep{rank1999estimation,IKEDA2010642}. Prior work has also focused on motion artifact recognition for MRI \citep{OKSUZ2019136} and CT images \citep{LOSSAU201968}, which mostly use supervised learning of convolutional neural networks (CNN). 
In the broader field of image quality assessment (IQA), and no-reference IQA (NR-IQA) in particular, reports on natural images are learning based and the early ones mainly use codebook \citep{6247789,QAC,7501619} or natural scene statistics (NSS) \citep{5756237,6172573,6272356,ILNIQE} for feature representation and extraction. Later, CNN based methods \citep{6909620,8063957,meon,8383698,Ying2020CVPR} dominate. 
These methods are trained to give one general quality score and/or assess the strength of multiple types of distortions respectively \citep{PONOMARENKO201557,live}. However, the types of distortions in camera images, except for noise, do not overlap with those in medical images, so these approaches can not be transferred directly to medical imaging. 
In the field of medical imaging, only a few NR-IQA works exploit artifact-specific  assessment or multi-task learning. \citet{ALI2021101900} uses bounding box based detection to outline regions of various artifacts for endoscopy videos. \citet{LIN2019101548} train a dual-task model that does quality-relevant object detection and binary classification simultaneously for ultrasound images. 
To the best of our knowledge, existing NR-IQA works for MRI only give a general quality class or score ambiguously defined within each work. These models are trained to perform a classification \citep{9090293,KUSTNER2018134,jmri.27649} or regression \citep{mrm.28201,ryai.2020190123} task, respectively.
 There are two types of constraints that come with the training dataset construction of the regression-based methods. Some \citep{8383698,6909936,QAC} require ground-truth images so that the difference between an image and the ground-truth can indicate its quality. Full-reference metrics like $\ell_1$-distance, FSIM \citep{5705575} and error maps are calculated and used as the training labels. The others use mean opinion score (MOS) as the training label. This requires multiple human raters to give a loosely defined subjective score for every single image in the dataset, which is labor-intensive. Therefore, datasets with MOS are usually small.  
\vspace{1.7mm}

In this work, we construct a training set that requires neither MOS nor reference images.
To do so, we first develop a deep learning based framework to perform diagnostic context and scan type dependent quality assessments from specific aspects of image imperfections. We then present a model developed under this framework: a dual-task CNN with divisive normalization (DN) nonlinearity. We then determine whether it can assess perceptual noise levels and detect motion corruption in MRI. The model is novel in multiple respects. First, no-reference perceptual scores are generated by human calibration which is more efficient and better defined for subjective labeling than conventional opinion scores. Second, we do inference with image rulers which is a visually interpretable and flexible way to accommodate radiologists' varying standards for MRI scan qualities. 

The remainder of this paper is organized as follows. Section \ref{S2} presents a framework, the methods used in the framework, and preparation of a dataset. Section \ref{S3} details the experimental results. Section \ref{S4} discusses the advantages, limitations, and possible extensions of this work. Section \ref{S5} summarizes the main ideas conveyed in this paper.

\section{Methods and material}\label{S2}
In this section, we first present a high-level multi-task framework for artifact-specific MRI quality assessment, followed by three subsections on methods we use for the perceptual noise assessment task, including a novel method of label generation and use of image rulers. Finally, we present the construction of a dataset, including physics-based simulation of noise and motion artifacts. 

\subsection{Multi-task framework}
Multi-task training is used for one model to learn to perform multiple different but related tasks. It has three advantages over using multiple independent single-task models. First, multi-task trained models generalize better. Second, for a main task, adding auxiliary tasks may boost the performance over a model trained only for that main task. Third, a multi-task model is more suitable for capturing correlated information. 

The first two advantages come from the sharing of layers by multiple tasks. When a network is trained on one task only, the optimization depends heavily on the most influential factors to the task and is more prone to overfitting. When trained on some related tasks in addition, the network learns additional helpful features that are ignored when focusing on one task. The third advantage comes naturally: single-task models trained independently ignore correlation between tasks (i.e. artifacts in particular), while a multi-task model has the potential of capturing this correlation to better extract features for each task. 

\begin{figure}[!t]
\centering
\includegraphics[width=0.9\linewidth]{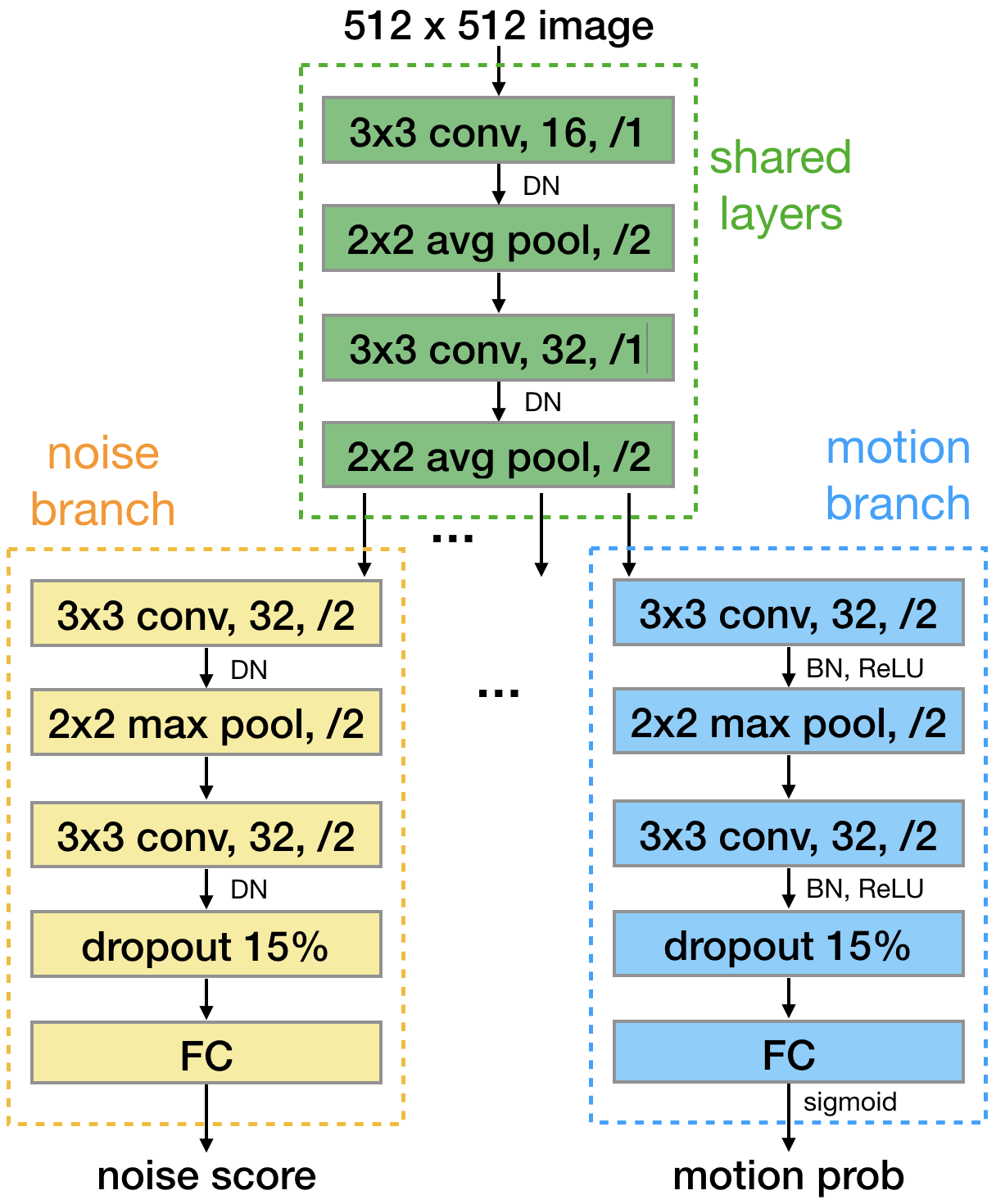}
\caption{Multi-task model architecture with noise and motion branches. More branches can be added.  }
\label{architect}
\end{figure}

We exploit a multi-task model where each branch of a branched CNN is responsible for assessing one type of artifact or one aspect of diagnostic quality. In this work, the model supports perceptual noise assessment and motion detection. Other types of artifacts such as fat suppression failure and magnetic field inhomogeneity can also be added. Fig. \ref{architect} shows the specific architecture used for models in this paper. In each block the convolution kernel or pooling size is followed by the number of feature maps and/or the stride number starting with `/'. DN or ReLU with batch normalization (BN) is used after the convolutional layers. The input to the network is a single-channel magnitude MRI of standard size. The output of the noise branch is a scalar score that can be interpreted by an image ruler. The output of the motion branch is a probability estimate that the input image is motion corrupted. 

The following methods in 2.2 to 2.4 are for the noise assessment task only. 

\subsection{Heuristic label generation} \label{sec:heuristic}
We use some baseline noise estimation methods, $Q$, to compute an approximate quality score $Q(X)$ for an training sample $x$ as the starting point for its training label $y$. This automatically generated heuristic score reduces the human labeling effort. Specifically, we use IEDD \citep{iedd}, a pre-trained MEON model \citep{meon}, and a SNR defined by \eqref{snr}-\eqref{asnr} as the function $Q$. Details of all three baseline methods are described below. 

IEDD is a deterministic method utilizing DCT coefficients. It does not involve learning, takes a single image and outputs an estimation of the WGN variance. In this case we have $Q(x)=IEDD(x)$. 

MEON is a CNN trained on MOS labeled natural image datasets. It takes a single image and outputs five scores for WGN contamination, JPEG2000 compression distortion, JPEG compression distortion, Gaussian blur, and general quality, respectively. We use only the outputs for WGN. In this case we have $Q(x)=MEON(x)^{[WGN]}$.  

We include a full reference metric and choose to approximately calculate the most standard measurement of noise level, SNR. Although we do not have noise-free images as the true reference for an accurate SNR, the following approximations should be informative enough for our purposes and comparable with the no-reference metrics.

For a set of five samples $x^{[1]-[5]}$ with decreasing amount of noise added to the original image $x^{[5]}$, we use the original image $x^{[5]}$ as the reference for calculating the SNR \eqref{snr} of all other simulated versions. The SNR is summed over pixels $x_p$ of $x$ and in dB. Then we assign an artificial SNR to the original version using \eqref{asnr}. Formally, when using SNR as the baseline method $Q(\cdot)$, we have
\begin{subnumcases}{Q(x^{[v]})=}
    10 \log_{10}{\frac{\sum_{p} \big|x_p^{[5]}\big|^2}
                  {\sum_{p} \big|x_p^{[5]}-x_p^{[v]}\big|^2}} 
        & for $v=1... \ 4$ \label{snr} 
    \\[7pt]
    \small Q(x^{[4]})+[Q(x^{[4]})-Q(x^{[3]})] & for \ $v= 5$. \label{asnr} 
\end{subnumcases}
\vspace{1mm}

\subsection{Label calibration} \label{sec:label}
Here we introduce a framework for generating visual quality label scores to address limitations of conventional methods used by NR-IQA works. 

The fastest way to obtain a score representing some visual quality of an image is to compare it with a gold standard image. A model can then be trained to learn a mapping between any given image and its distance from the perfect image so that in the inference phase IQA is done on individual test images without a reference. The label scores in this case are automatically calculated using reliable FR-IQA metrics or measures of distributions, so no human label is involved. 
This approach is not suitable for diagnostic IQA tasks because of two reasons. First, some noise and motion artifacts are present in all MR images, so no true ground-truth image can be used even for training. Second, radiologists' perception of image imperfections are important as they implicitly incorporate diagnostic goals. Hence, human input is essential for training.

\begin{figure}[!t]
v= 0 \hspace{5mm} 1 \hspace{11mm} 2 \hspace{11mm} 3 \hspace{12mm} 4 \hspace{12mm} 5 \hspace{6.5mm} 6\\
\includegraphics[width=0.99\linewidth]{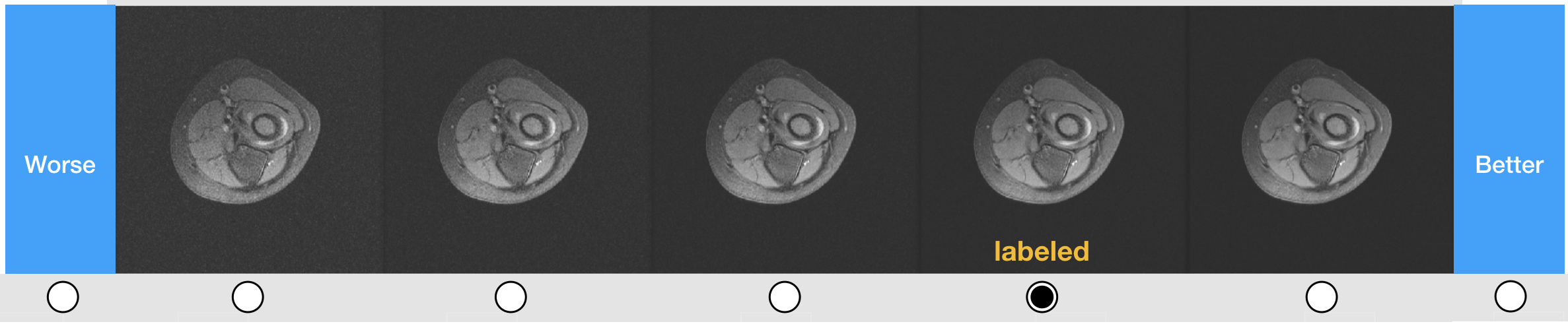}
\caption{Labeling of one set of images. The rater is asked to choose one version with the minimum acceptable quality. The rater can choose from one of the given $m_t=5$ images and anything outside ($v=0$ or $6$) of the range covered by given samples. In this example, the 4th image is chosen so $h_i=4$.}
\label{gui}
\end{figure}

The most straightforward way of obtaining visual scores is to directly ask human raters to give that score. A rating rubric is defined by some literal description. There is subjective variation when linking the literal description to the visual appearance, so a large sample size of human labels is needed to marginalize this variation. This is the commonly used MOS. Thus, multiple human labels are required for one training sample. 

In this paper we will show  we do not have to train models on the ultimate goal (i.e. output MOS) for them to serve an equivalent purpose in inference. We can learn from  relevant tasks that depend on the same features used for the ultimate task. \citet{7934456,9191231} have shown that merely by learning from binary comparisons of a certain quality between pairs of two images, the network can provide a score that correlates well with that quality of a single image. Human labeling by picking a better image between two is better-defined and easier in this approach than that in MOS. At least one human label is required for every two training samples. 

Our approach is to learn from a human calibrated score, which is more closely related to the MOS than pairwise comparison results. Our training dataset consists of $N$ sets of images. Each set consists of $m_t$ versions ($v=1,...,m_t$) of a unique 2D slice $x_i$. The amount of noise in the images decreases linearly from $v=1$ to $v=m_t$. We first generate a heuristic score $y=Q(x)$ for each image $x$ in the training set as described in section \ref{sec:heuristic}. Then we ask an experienced radiologist to pick the version $h_i$ in each set with the minimum acceptable perceptual noise level for this unique slice. This selection is \textit{context-dependent} and reflects the rater's perception of noise in these specific slices. Fig. \ref{gui} shows one set of images in the rating interface. Unlike for MOS, our rater is not required to strictly follow a given rating standard defined in words. The standard for ``minimum acceptable'' can be interpreted by the rater as long as he or she is consistent through rating the whole training set. 

\begin{figure}[!t]
\includegraphics[width=0.97\linewidth]{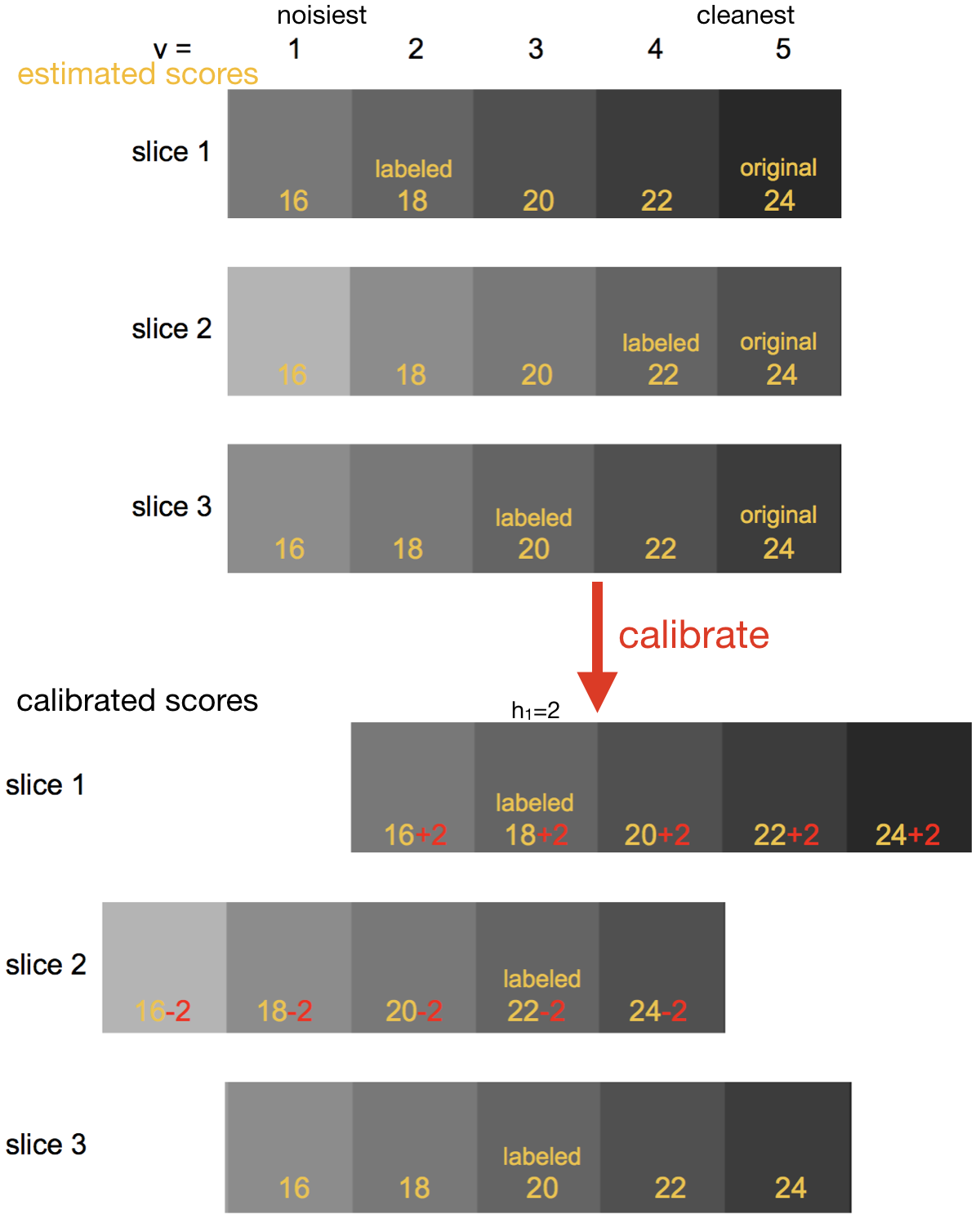}
\caption{Demonstration of the calibration process with a simplified example. Each block represents an image sample in the training set. Each of the three unique slices has five versions. Brightness of the blocks represents the perceptual noise level. The estimated scores partially capture the noise level but have error: blocks with the same estimated score do not have the same brightness. After a rater labels out blocks with a same brightness, their scores are calibrated to the mean (20) and scores of their other versions shifted accordingly.}
\label{calib}
\end{figure}

We use the labeled versions as the calibration point because they have similar perceptual quality by definition of the labeling task.  Therefore, scores $y_i^{[v]}$ of all images in a set are adjusted, following \eqref{e3}, according to the deviation of the $y_i^{[h_i]}$ in this set from their mean, $\mu_h$ in \eqref{e2}. Coefficient $\eta$ controls the adjustment amount, i.e., the calibration strength. 
%
%
The labeled images have similar qualities by definition of the labeling task. However, their qualities cannot be exactly the same, because there is a finite number of versions for the raters to select from. Therefore, we include $0<\eta\leq 1$ to allow a soft shift toward the calibration point instead of forcing the score of all labeled images to be the same. Fig. \ref{calib} illustrates the intuition for the calibration process in a simplified setting where $\eta=1$. By enforcing the labeled images to have similar scores and adjusting the other versions accordingly, this calibration process reduces the bias between perceptual noise level and the baseline quantitative estimation. The steps are formally defined as follows, where we start from the heuristic score $y$ and arrive at the calibrated score $\hat{y}$. 

\begin{equation}
\label{e1} 
y^{[v]}_i=Q(x^{[v]}_i),\\
\end{equation}
\begin{equation}
  \label{e2}  
\mu_h=\frac{1}{N} \sum^N_{i=1}y_i^{[h_i]},\\
 \end{equation}
\begin{equation}
\hat{y}^{[v]}_i =
    \begin{cases}
    y^{[v]}_i+\eta\ \big(\mu_h-2y_i^{[1]}+y_i^{[2]}\big) & \text{for }\ h_i= 0\\[7pt]
      y^{[v]}_i+\eta\ \big(\mu_h-y_i^{[h_i]}\big) & \text{for }\ h_i= 1 ...\ m_t \\[7pt]
      y^{[v]}_i+\eta\ \big(\mu_h-2y_i^{[m_t]}+y_i^{[m_t-1]}\big) & \text{for }\ h_i= m_t+1.
    \end{cases}
    \vspace{1mm}
    \label{e3}
\end{equation}
The first and last cases in \eqref{e3} follow the same idea of the middle case, where we estimate a virtual $y_i^{[h_i]}=y_i^{[m_t+1]}=y_i^{[m_t]}+(y_i^{[m_t]}-y_i^{[m_t-1]})$.

We have introduced a way of defining perceptual scores for training with more information than binary comparisons, or less labeling effort than conventional methods. By utilizing an automatic heuristic estimation followed by human-aided calibration, perceptual score labels for multiple (more than five) samples can be generated with only one human label.

\subsection{Image ruler} \label{sec:eff}
An image ruler is a sequence of $m_r$ versions of a unique slice $x_r$ with decreasing amount of noise. The worst ($x_r^0$) and best ($x_r^{[m_t-1]}$) versions in the image ruler are chosen to cover the range of qualities seen in clinical practice. A sample image ruler is shown in Fig. \ref{inference} and all nine image rulers used in this work are shown in Appendix A. 

The inference flow using the image ruler is illustrated in Fig. \ref{inference}. For each test image, an image ruler of a scan of the same body part with similar image contrast is used. First, all versions in the image ruler are passed into the CNN, $D(\cdot)$, whose raw outputs form a vector $\mathbf{S}_{ruler}\in\mathbb{R}^{m_t}$. Then the same CNN's outputs on test images $x_{test}$ are compared to elements in $\mathbf{S}_{ruler}$ to get a \textit{ruler score} for each test image. A ruler score $RS$ is formally defined as follows.
\begin{equation}
    RS(x_{test})=\arg\!\min_v\big|\ D(x_r^{[v]})-D(x_{test})\ \big|.
    \label{rscore}
\end{equation}
Radiologists can indicate the appropriate quality standard specifically for each type of scan by choosing a pass-fail threshold in its image ruler. This threshold can be at the mid-point between two versions ($t_a<t_b$) or at one version ($t_a=t_b$) in the image ruler. Then we determine whether a test image passes the quality check ($PF=1$) or fails to meet the standard ($PF=0$) according to: 
\begin{equation}
    PF(x_{test})=\mathbbm{1}\big\{D(x_{test})\geq \frac{D(x_r^{[t_a]})+D(x_r^{[t_b]})}{2}\big\}. 
    \label{pf}
\end{equation}

\begin{figure}[!t]
\includegraphics[width=0.98\linewidth]{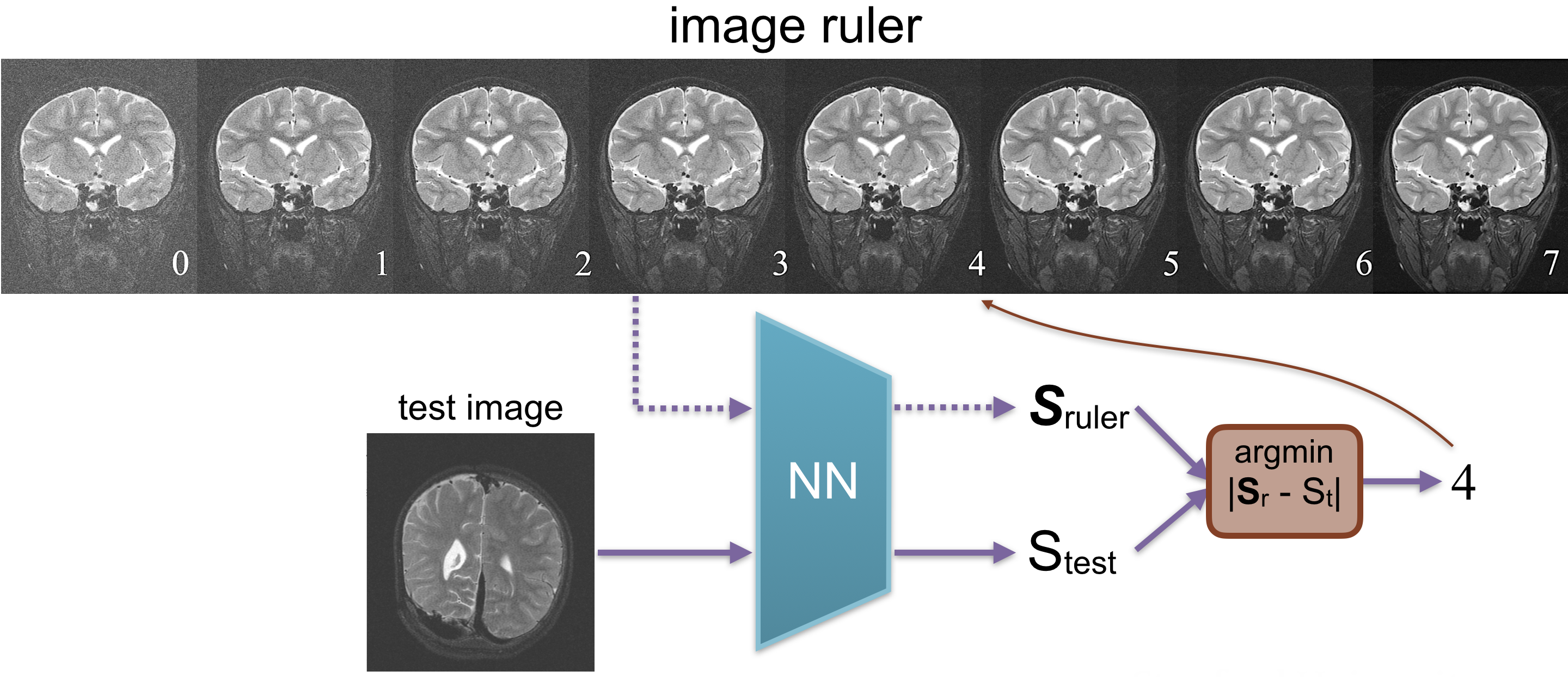}
\caption{Inference pipeline with an image ruler. First a vector $\mathbf{S}_{ruler}$ is obtained by passing all eight images in the image ruler to the trained NN (i.e. $D$ in equations). Then a scalar ${S}_{test}$ is obtained by passing a test image to the same NN. Finally a ruler score of 4 is given by finding the entry in $\mathbf{S}_{ruler}$ closest to ${S}_{test}$, indicating that the test image has a perceptual noise level as the \#4 image in the ruler.}
    \label{inference}
\end{figure}

The image ruler provides an intuitive and visually concrete way of interpreting the raw real number output from the CNN. The CNN is trained on label scores that are approximately the same for images with similar perceptual noise levels, and the ruler score is determined by the proximity of raw scores for test and ruler images. Therefore, given a ruler score, one can infer that the test image has a quality as shown in the corresponding version in the image ruler. Conversely, the image ruler also serves as a visual interface for radiologists to pick critical quality threshold. Compared to the basic approach of classification models with fixed quality levels vaguely defined from good to bad, the approach of scalar output combined with image ruler defined thresholds provides flexible and visually defined levels of quality. 

\begin{figure*}[!t]
\includegraphics[width=0.95\linewidth]{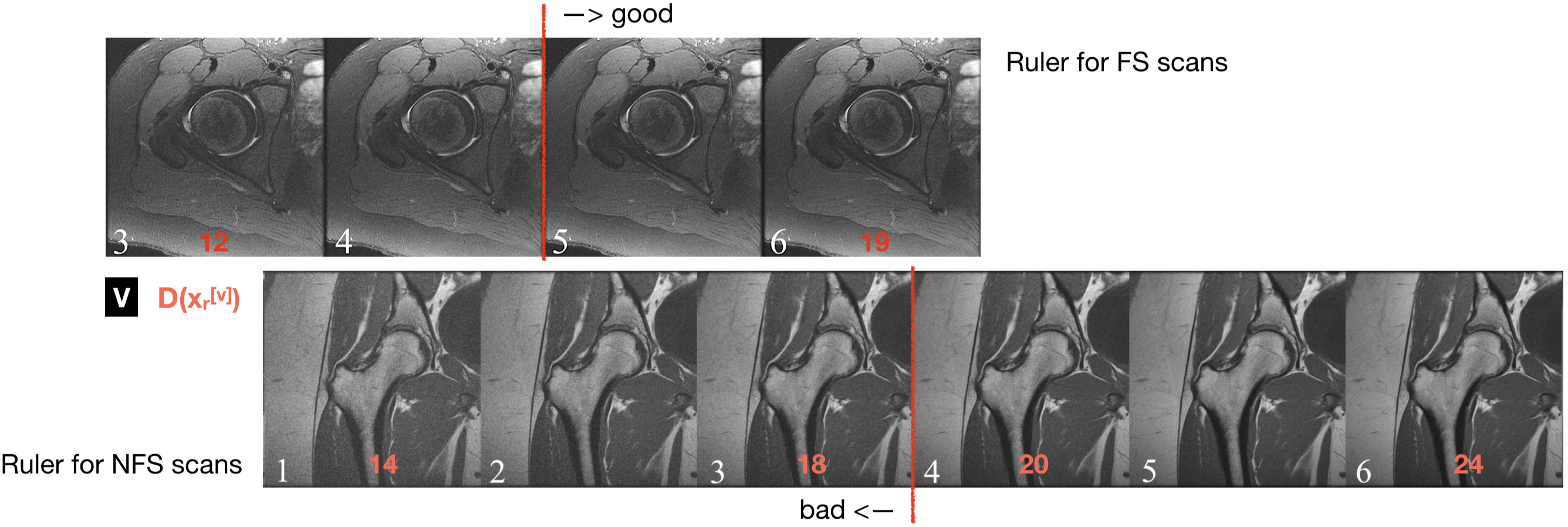}
\caption{Comparison of two partial hip image rulers for fat suppressed and  non-fat suppressed scans, respectively. Their positions are roughly aligned on the perceptual noise level. The red vertical lines mark the thresholds of determining if an image is acceptable for diagnostic purposes. The numbers in red represent the model's raw outputs. We can see a good image, \#6, from fat suppressed is only barely acceptable by the standard of  non-fat suppressed. Furthermore, a single standard used for both would yield more false positives (good) for non-fat suppressed or false negatives (bad) for fat suppressed. }
\label{aruler}
\end{figure*}

Image ruler defined thresholds are helpful also because different standards are needed for different scans or diagnostic tasks. For example, fat suppressed images are noisier than non-fat suppressed images on average. However,  the acceptable noise level is higher for fat suppressed images. Fig. \ref{aruler} shows two partial hip image rulers roughly aligned for their perceptual noise level. A ruler score in the fat suppressed ruler corresponds to a lower score in the non-fat suppressed ruler. Assuming equal diagnostic tasks, the selected threshold on the fat suppressed rulers should correspond to a worse raw score from the CNN. 
Therefore, for the task of determining whether an image is adequate, scan-dependent thresholds determined by image rulers should give better accuracy than an optimal fixed threshold. 

In theory one could pick two thresholds from one ruler that are numerically equivalent to one from each ruler. However, radiologists perceive fat suppressed and  non-fat suppressed images differently.  Thus, an image quality comparison between fat suppressed and  non-fat suppressed images, or between hip and brain images, is ill-defined. Having separate image rulers for different scans also makes it easier for a model to perform the matching and give accurate ruler scores. 

We aim to avoid effects from the irrelevant features (i.e. contrast, object shape, etc.) of the image content on the estimated perceptual noise level by the use of image ruler. The calibration step for training reduces part of the effect of image content on the baseline noise estimation. With scan-specific image rulers, we can further reduce the undesired effects in the inference phase. Because the ruler scores $RS$ is obtained by comparing the raw output of the same NN on the test and corresponding ruler images, the content introduced bias is canceled out. And the more similar the content of the ruler is to the test image, the more accurate ruler score $RS$ is obtained, where the extreme case is when the slice used to generate the image ruler is the same as the test slice. 

Thus far we have shown how we use image rulers to determine whether an image is acceptable for diagnostic purposes in terms of perceptual noise level and to produce a ruler score that can be interpreted visually. We have also explained that the use of image rulers in inference is helpful for both human understanding of the model output and the model's performance.

\subsection{Divisive normalization}
We use divisive normalization (DN) \citep{DNT} in place of the commonly used ReLU with batch normalization (BN) \citep{3045118.3045167} or layer normalization (LN) \citep{Ba2016LayerN} as the nonlinearity after convolutional layers for the noise assessment task. 

The divisive normalization transform is a joint transformation that is highly effective in Gaussianizing local patches of natural images \citep{gdn}. It is a form of local gain control modeling the nonlinear properties of cortical neurons \citep{Carandini2013}. \citet{4799311} have shown that statistics of DN transformation domain representations of images are sensitive to distortions like Gaussain noise, Gaussain blurring and JPEG compression, and it is effective in assessing natural image quality. The equation below defines the DN transformed $a_i$ for each spatial point $z_i$ in each channel $j$ of the convolution output: 
\begin{equation}
    a_i=\frac{z_i}{\big(\ \beta_i+\sum^C_{j=1} \gamma_{ij}^{\textcolor{white}{1}}\ z_j^2\ \big)^{\frac{1}{2}}} \ .
\label{gdn}
\end{equation}
The summation in the denominator is across $C$ channels corresponding to the $C$ feature maps of the preceding convolutional layer. Bias $\beta\in \mathbb{R}^M$ and weight $\gamma\in \mathbb{R}^{M\times C}$ are trainable parameters of the DN activation layer, where $M$ is the number of elements in one 2D channel of the input to this layer. Unlike affine BN and LN, DN itself is a nonlinear transform with a high degree of nonlinearity. 
We find DN suitable for the Gaussian distributed noise artifact but not for the structured motion artifact. We empirically find that DN alone works better for the noise assessment task than the commonly used combination of nonlinearity and normalization. A performance comparison between DN, ReLU with BN, and ReLU with LN is included in Appendix B.

\subsection{Dataset construction}
\noindent \textbf{Standard dataset.} With IRB approval and informed consent, we collected raw scanner data from clinical scans performed on two 3T GE MR750 whole-body MR scanners and various receiver coils whose coil numbers range from 8 to 32. 
The raw data provides k-space measurements with acquisition details, and other scan descriptions and parameters. 
In total, there are 9778 unique slices from 635 subjects. All slices are interpolated to the size of $512 \times 512$ following the convention for MRI. All images are from fully-sampled 2D fast spin echo (FSE) scans that are T1, T2, FLAIR, or proton density weighted, with or without fat suppression, and are from pediatric subjects. 
$46\%$ of the scans utilized partial Fourier with a various number of extra phase encoding lines past the center of k-space. The scans were acquired with an echo train length ranging from 1 to 23.
The anatomic regions in this dataset are: brain, elbow, hip, spine, knee, ankle, and foot, where spine images are only presented in the test sets. There are 470 slices in the test set for noise and 295 slices in the test set for motion. 

\noindent \textbf{OOD dataset.} We include a dataset randomly drawn from a clinical database as the out-of-distribution (OOD) dataset to test the generalizability of our model. This dataset consists of series of reconstructed $512 \times 512$ images in DICOM format with scan descriptions and parameters.  
In total, there are 180 unique slices from 45 subjects. All data are from 2D FSE scans and all anatomic regions presented in the standard dataset are included. Data in this dataset differs from that in the standard dataset in terms of patient age, scanner hardware, reconstruction method, etc. There are, with overlap, 140 slices in the test set for noise and 120 slices in the test set for motion.

As described below, we simulate noise and rigid motion in MRI, for the training of our model, based on standard physical models of MRI acquisition. 

\subsubsection{Noise simulation}
For each unique slice, we generate $m_t-1$ noise injected versions to include in the training set with the original slice. First we add complex zero-mean white Gaussian noise (WGN) to the original multi-coil k-space data $K$, according to the physical noise model of MRI \citep{mrm.1910360327}. WGN with the same variance is added to the real and imaginary k-space samples for of each coil. With $m_t=5$, four decreasing variances are used for $v=1$ to 4 in $x^{[v]}$ such that the SNR in dB increases linearly. Then the noise injected k-space and the original k-space are both reconstructed by the sum-of-square reconstruction
\begin{equation}
    x(p_x,p_y)=\sqrt{\sum_{i=1}^{N_c} \big|I_i(p_x,p_y) \big|^2} \ ,
    \label{sos}
\end{equation}
where $I_i=\mathcal{F}^\text{-1}\{K_i\}$ is the $i$th coil image from the inverse 2D Fourier transform of the $i$th coil k-space, $(p_x,p_y)$ is the spatial location of one pixel in the image, and $N_c$ is the number of coils.
 For the image rulers with $m_r=8$, we simulate seven noisier versions of the original slices ($v=7$) where the WGN variance decreases from $v=0$ to 6.

To calibrate training labels for the noise dataset, we collect 769 human labels, at least one labeled slice for each subject. The label for each slice is set to that of its closest labeled slice.  

\subsubsection{Motion simulation}
\begin{figure}[!t]
\centering
\includegraphics[width=0.97\linewidth]{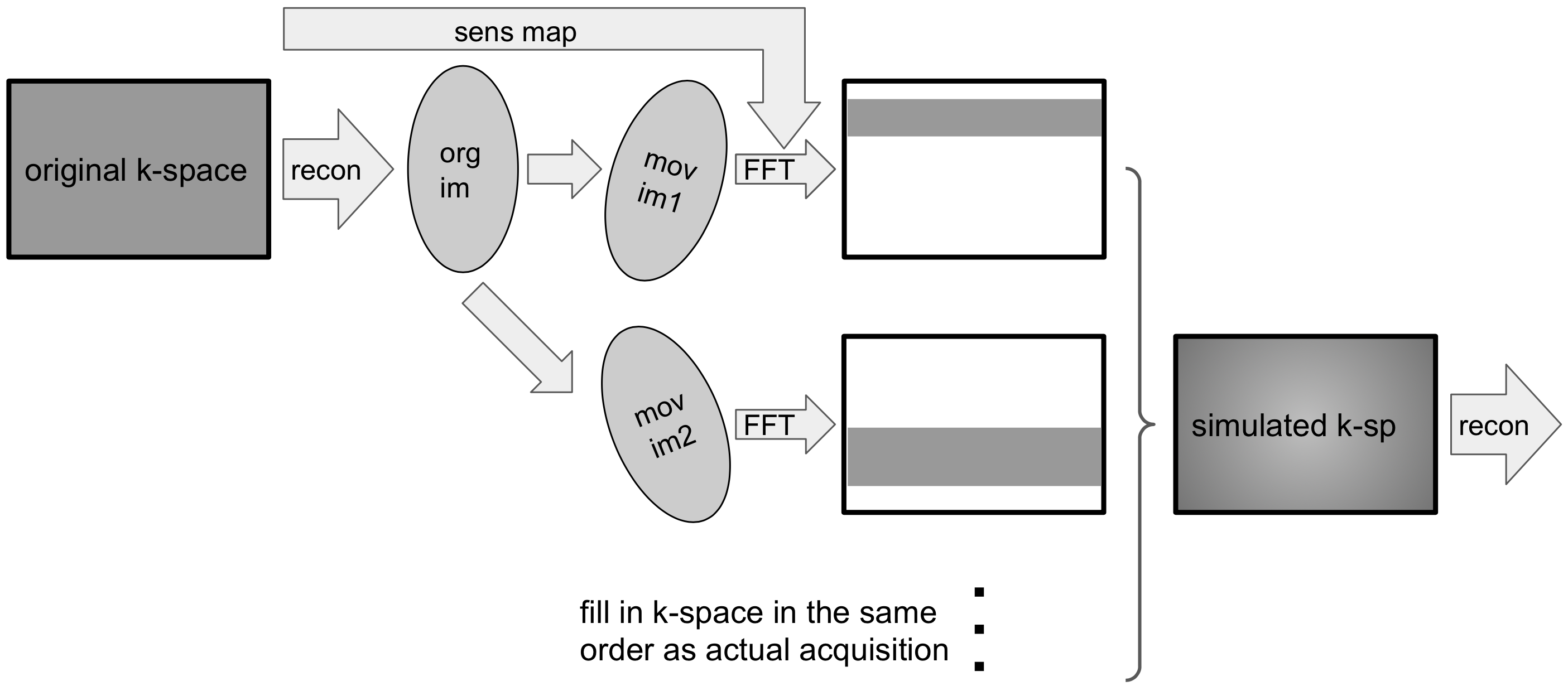}
\caption{Motion simulation pipeline. First the original image and coil-sensitivity maps are obtained from the original k-space measurements. Then the original image is moved to multiple positions, simulating the movements of the object being scanned. Portions of the simulated k-space are filled, by ``scanning'' the moved images, following the actual acquisition process. Finally the motion simulated image is reconstructed from the simulated k-space.}
\label{motionsim}
\end{figure}

For each unique slice, we generate one motion injected version to add to the training set. First, we estimate the coil sensitivity maps $s$ with the ESPIRiT calibration method \citep{mrm.24751}. Then the multi-coil reconstruction \eqref{mc} gives a complex-valued image $x_o$ whose pixels are given by:
\begin{equation}
    x_o(p_x,p_y)=\frac{\sum_{i=1}^{N_c} I_i(p_x,p_y) s_i^*(p_x,p_y)} {\sqrt{\sum_{i=1}^{N_c}|s_i(p_x,p_y)|^2}}, 
\label{mc}
\end{equation}
where $^*$ stands for the complex conjugate. $x_o$ is rotated and shifted (with spline interpolation) to $x_m$, then transformed back to multi-coil k-space $\hat{K}$ using the forward model of MRI, for which \eqref{mc} is the least square solution. The $i$th coil in $\hat{K}$ is given by:
\begin{equation}
    \hat{K}_i=\mathcal{F}\{x_m\odot s_i\},
\end{equation}
where $\odot$ is the element-wise multiplication. Part of $\hat{K}$ is filled into the final simulated k-space, according to the actual order of acquisition of this specific scan. The process is repeated with different motion positions until all k-space is filled, as illustrated in Fig. \ref{motionsim}. The final simulated k-space is reconstructed with \eqref{sos} to give the motion simulated image.
The number of moves is randomly chosen from two to four. The amounts of rotation and shift are independently randomly chosen within 1 degree and 3 pixels from the last position, respectively. The time between two moves is chosen randomly. Samples of simulated images are included in Appendix C.

Training labels for the motion dataset are automatically generated.  All motion injected images labeled as fail and all original ones labeled as pass. Although this does not guarantee accurate labels, we rely on our model's robustness to labeling errors.

\subsubsection{Test set labeling}
For the training set, one rater labels by making a selection that is not directly for the final tasks. For the test set, other two raters label directly for the final tasks, which are giving a ruler score for noise assessment and determining if an image is adequate in terms of noise and motion.   

We ask two radiologists to independently give a ruler score for each image in the noise test set. Specifically, a test image and its corresponding image ruler are presented to the rater simultaneously. The rater is asked to pick from one of the eight versions in the image ruler that has the most similar perceptual noise level as the test image. The raters are also asked to indicate whether this test image is acceptable for diagnostic purposes. 
 
We ask one radiologist to pick out images in the motion set that are motion corrupted, which are labeled as fail, while the rest are labeled as pass. We only have one rater to label the motion dataset because it is a more trivial binary task and we do not expect as significant of inter-rater difference as for the noise task. 
 
\begin{table*}[!t]
  \caption{\label{tab1}Binary classification accuracy and mean absolute ruler score error of the single-branch noise assessment network (NN-) trained with uncalibrated SNR, IEDD, MEON labels and their calibrated (c*) versions, respectively. The best two results are in bold. }
  \vspace{1.7mm}
  \centering
  \begin{tabular}{|p{1.7cm}||p{2cm}|p{2cm}|p{2cm}|p{2cm}|p{2cm}|p{2cm}|}
    \hline
    Metric & NN-SNR & NN-cSNR & NN-IEDD & NN-cIEDD & NN-MEON & NN-cMEON \\
    \hline
    Accuracy & 79.79\%  & 83.19\%  & 82.34\% & \textbf{89.57\%} & 85.96\% & \textbf{91.06\%} \\
    Score error & 1.340 & 1.279 & 1.298  & \textbf{1.098}  & 1.189 & \textbf{1.066} \\
    \hline
\end{tabular}
\label{T1}
\end{table*}

\subsection{Network training}
Root mean square error and binary cross-entropy are used as the training losses for the noise and motion tasks, respectively. Parameters of the two branches in Fig. \ref{architect} are optimized alternatively every mini-batch through a stochastic gradient descent algorithm \citep{adam}. The shared layers are updated every mini-batch. Formally, let $D_n$ and $D_m$ be the neural networks that consist of the shared layers followed by the noise and motion branches, respectively. Let $x$ be the input image and $\hat{y}$ or  $\Tilde{y}$ be its label. Let $p_n$ and $p_m$ be the distribution of training images for noise and motion. The training losses for noise and motion task are defined by \eqref{rmse} and \eqref{bce}, respectively.

\begin{equation}
     \mathcal{L}_{\text{noise}}=\sqrt{\mathbb{E}_{x \sim p_n} |\hat{y}-D_n(x)|^2}, 
\label{rmse}
\end{equation}
\begin{equation}
    \mathcal{L}_{\text{motion}}=-\mathbb{E}_{x \sim p_m} \Tilde{y}[log(D_m(x))]+(1-\Tilde{y})[log(1-D_m(x))].
\label{bce}
\end{equation}

Each mini-batch for optimizing \eqref{rmse} consists of two sets of five $x$: four noise injected versions and the original version of a unique slice. Each mini-batch for optimizing \eqref{bce} consists of five pairs of $x$: the motion injected and the original version of a unique slice. $\hat{y}$ is obtained as described in section \ref{sec:label} and \eqref{e1}$\sim$\eqref{e3}. $\Tilde{y}$ is set to 0 (motion) for all motion injected versions and and 1 (no motion) for all original versions.  
More training details including the optimization hyper-parameters are included in Appendix C.

\section{Results}\label{S3}
In this section we compare training the noise assessment model with labels before and after the proposed calibration, and inference with and without image rulers. Then we compare our method with six existing NR-IQA methods and human performance. Finally, we compare the single-task and dual-task models on their generalizability with the OOD dataset. In all experiments, we have the number of versions for each unique slice in the training set $m_t=5$, and the number of versions in each image ruler $m_r=8$ .

\subsection{Effectiveness of label calibration}
We test the effect of label calibration on three baseline noise estimation methods $Q$: IEDD \citep{iedd}, a pre-trained MEON model \citep{meon}, and a SNR defined by \eqref{snr}-\eqref{asnr}.

We search for the best calibration strength $\eta$ in \eqref{e3} using the MEON baseline from [0.5, 0.6, 0.7, 0.85, 1] and find $\eta=0.85$ to work best. This is then used for all experiments. The human labels $h_i$'s collected for \eqref{e3} are mostly between 2 and 6. More training details are included in Appendix C.

Test results for the model trained on six sets of labels are shown in Table \ref{T1}. We provide three pairs of comparisons on the model's performance on estimating a ruler score, evaluated by the mean absolute score error, and determining whether an image is adequate, evaluated by the binary classification accuracy. The CNN, referred to as NN in the table, used here is the single-task model with the noise branch only. For example, the first pair NN-SNR and NN-cSNR stand for the CNN trained with raw and calibrated SNR labels, respectively. 

Three pairs of comparisons in Table \ref{T1} consistently show that the proposed calibration step significantly improves model performance, regardless of the type of baseline estimation. Comparison across the types of baseline shows that labels based on MEON best serve the training of our model. From this point on, we only consider the best performing version, NN-cMEON, for our noise assessment branch.

\begin{table*} [!t]
\renewcommand\arraystretch{1.2}
  \caption{Binary classification accuracy, absolute error and Krippendorff's alpha for the ruler score from six existing methods, our best performing model (NN-cMEON), and human professionals. Comparison of results from an optimal single-value threshold and thresholds defined by two or nine content-dependent image rulers. The best two results are in bold. Marginal errors denote the 95\% confidence interval.  }
  \vspace{1.7mm}
  \small
  \centering
  \begin{tabular}{ |p{1.6cm}|p{1.9cm}||p{1.1cm}|p{1.6cm}|p{1.4cm}|p{1.2cm}|p{1.cm}|p{1.cm}|p{1.7cm}|p{1.5cm}|}
    \hline
    Metric & Threshold type & QAC  &Chen (2015) &   Liu (2012) & ILNIQE &IEDD & MEON   & NN-cMEON & Radiologists \\
    \hline 
     & Single best &  61.70\% & 62.77\% & 68.72\% & 70.85\% & 69.79\% & 76.38\% & 85.11\% &   \\
    Accuracy & 2 ruler defined & 66.38\% & 71.06\% & 74.04\% & 75.74\% & 78.94\% &  81.06\% & 88.72\% & \textbf{87.23\%} \\
    & 9 ruler defined &  68.09\% & 72.13\% & 75.11\% & 78.94\% & 80.85\% & 84.47\% & \textbf{91.06\%} &  \\
    \hline
    Score error & 2 ruler defined & 2.264 & 1.813 & 1.611 & 1.681 & 1.638 & 1.309 & 1.126 ~ & -- \\
    & 9 ruler defined & 2.213  & 1.770 & 1.553 & 1.651 & 1.432 & 1.209 & \textbf{1.066} & \textbf{0.987} \\ 
    \hline
    Krippen- dorff's alpha & 9 ruler defined & 0.263$\pm$  0.075  & 0.528$\pm$  0.061 & 0.573$\pm$  0.056 & 0.543$\pm$ 0.059 & 0.615$\pm$ 0.054 & 0.705$\pm$  0.047 & \textbf{0.762}$\pm$ ~ 0.039 & \textbf{0.783}$\pm$  0.037 \\ 
    \hline
  \end{tabular}
\label{t2}
\end{table*}

\subsection{Effectiveness of the image ruler}
Inference with image rulers gives a concretely interpretable ruler score, in contrast to standard regression methods. It also enables customizable thresholds for passing and failing images, also in contrast to standard classification methods. 
The basic way to convert a regression output to classes is to find a raw score threshold between classes that gives the most preferred classification performance on an evaluation set, analogous to selecting the best operating point on the receiver operating characteristic (ROC) curve. 

We compare using content-specific thresholds extracted from image rulers to the single-best fixed threshold for the task of determining if an image looks too noisy. 
As defined in \eqref{pf}, a ruler-defined threshold is $\frac{D(x_r^{[t_a]})+D(x_r^{[t_b]})}{2}$, where $x_r^{[t_a]}$ and $x_r^{[t_b]}$ are one or two of the eight images in one of nine image rulers. We do a linear search of 100 values between the lowest and highest $D(x_r^{[v]})$ across all image rulers for the single best threshold, where $v$ ranges from 0 to 7 and $t_a$ and $t_b$ range from 2 to 5. Here we use accuracy as the measurement of classification performance, so the single best threshold is the value that results in the highest accuracy on the same evaluation set used to validate the training of our CNN. 

The second and fourth row in Table \ref{t2} list the classification accuracy when using the single best threshold and the complete set of nine ruler defined thresholds, respectively. For all seven methods that output a scalar score, classification results converted from image ruler defined thresholds have an accuracy on average 7.9\% higher and at least 6.0\% higher than those from the single best threshold. This is mainly due to the fact that radiologists have different standards on the noise level when classifying each of the test images as good or bad, and this difference is captured by the scan-specific image rulers. 

We also compare using nine image rulers to using two image rulers. The nine rulers account for two fat suppression settings and the difference in anatomic regions. The two rulers account for the two fat suppression settings only. They are from a knee and an elbow scan, respectively. Therefore with two image rulers, most test images are matched to a ruler of the same fat suppression setting but a different anatomic region.  

The comparison between using two and nine rulers, listed in the last four rows in Table \ref{t2}, demonstrates the effect of having image rulers that are more content-specific. With the two image rulers and being specific for fat suppression, the performance significantly improves from using a single threshold. Then the set of nine rulers that is additionally specific on anatomic region further improves the performance from that of the two rulers. This observation supports the hypothesis in section \ref{sec:eff}.
Effects of irrelevant content are reduced and more accurate ruler scores are obtained as we go from non-content-dependent to fat-suppression-dependent, then from non-anatomy-specific to anatomy-specific rulers. 

We then experiment with extending the support to types of scans not included in the training by merely adding corresponding image rulers in the inference. Specifically, we train our model on a training set without any spine image, generate an image ruler with one spine images, then use the trained model to test on 42 spine images and convert the raw inference outputs to ruler scores and bad/good classes using this new spine image ruler. 

The classification accuracy and ruler score error on the spine images are 88.10\% and 1.167, respectively, 3\% and 0.1 worse than the model's performance on seen scans. This slightly degraded performance is still clinically valuable.
Therefore, for unseen types of scans, if further training of the model is not applicable, we can simply generate an image ruler for it. This result also supports our hypothesis illustrated in section \ref{sec:eff}, that the use of a similar-looking image ruler contributes to getting an accurate ruler score. 

We have shown that using content-dependent image rulers to classify images is superior to the fixed threshold approaches. Moreover, additional image rulers can be incorporated without retraining the model.

\subsection{Performance}
The two previous subsections have shown that training with calibrated MEON scores and inference with nine image rulers give the best-performing model. Now we compare our best performing model with six existing noise estimation or NR-IQA methods and radiologists. We then provide more detailed evaluations on the performance of our model. 

We pick six existing methods covering all major types of approaches that output a scalar for an image input, from statistical noise estimation methods to codebook, NSS, and deep learning based NR-IQA methods. 
IEDD \citep{iedd} and works by \citet{Chen} and \citet{Liu} estimate the variance of noise in images without training data. represent statistical noise estimation. 
QAC \citep{QAC} uses codebook and quality-aware clusters of patch features to predict the image quality. Its training uses the full-reference metric FSIM so reference images are needed. 
ILNIQE \citep{ILNIQE} uses NSS and multivariate Gaussian model to compare test images to a set of pristine images. The authors train their model on 90 pristine images. We use an ILNIQE model retrained on 100 relatively clean images from our dataset.
MEON \citep{meon} uses deep learning and a CNN trained with MOS to predict five quality scores corresponding to five aspects of image quality. We use the best pre-trained model provided by the authors and only one of the scores corresponding to noise. 

Table \ref{t2} lists the binary classification accuracy and the absolute ruler scores error of all methods and the human performance. It also includes the Kripfpendorff's alpha coefficient \citep{KsA} reflecting the level of agreement of the ruler scores between the model and the label. Here we focus on the comparison when using the better nine-ruler-defined threshold, although our model outperforms other methods in all settings. 
On the test set of 470 images, our model achieves an accuracy of 91.06\%, on average 14.5\% higher and at least 6.6\% higher than the other six methods. Our model achieves a ruler score error of 1.066, out of 7, on average 0.55 lower and at least 0.14 lower than the other six methods.

\begin{figure}[t]
\centering
\includegraphics[width=0.95\linewidth]{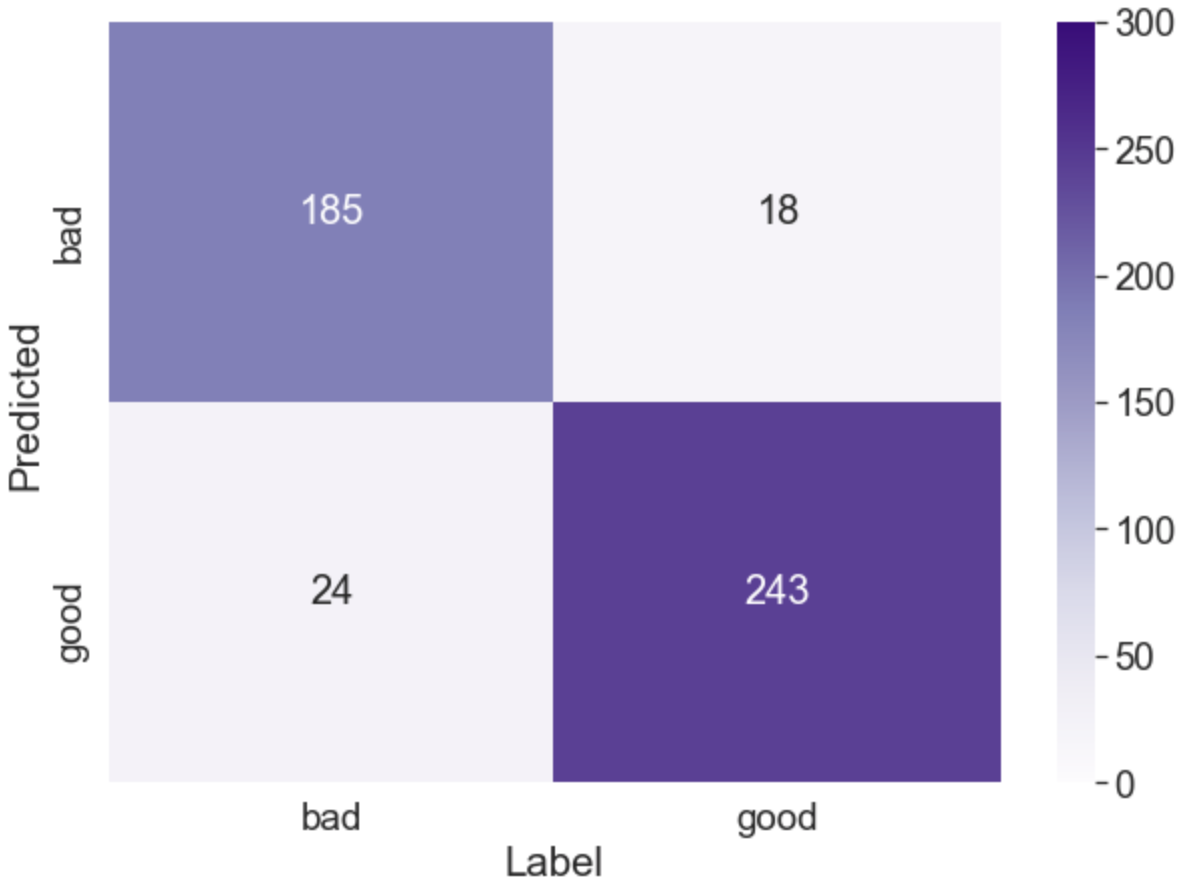}
\caption{Confusion matrix for the task of determining whether an image is good enough or has too much artifact (bad). Predictions are from the single-branch noise model on the standard test set, with an accuracy of 91.06\%. }
\label{mat2}
\end{figure}

The results for radiologists come from the difference between labels given by two radiologists rating the test set. Our model is 0.085 worse than radiologists in terms of the ruler score, but it exceeds human performance in the binary classification task. No other method reaches near-human performance. 

Fig. \ref{mat2} is the confusion matrix for our noise assessment model on the binary classification task of determining whether an image is adequate. The classification thresholds chosen in the nine image rulers lie between the 2nd to the 5th version (more detail in Appendix A). We have roughly balanced true positive and true negative rates. 
   
\begin{figure}[!t]
\centering
\includegraphics[width=0.97\linewidth]{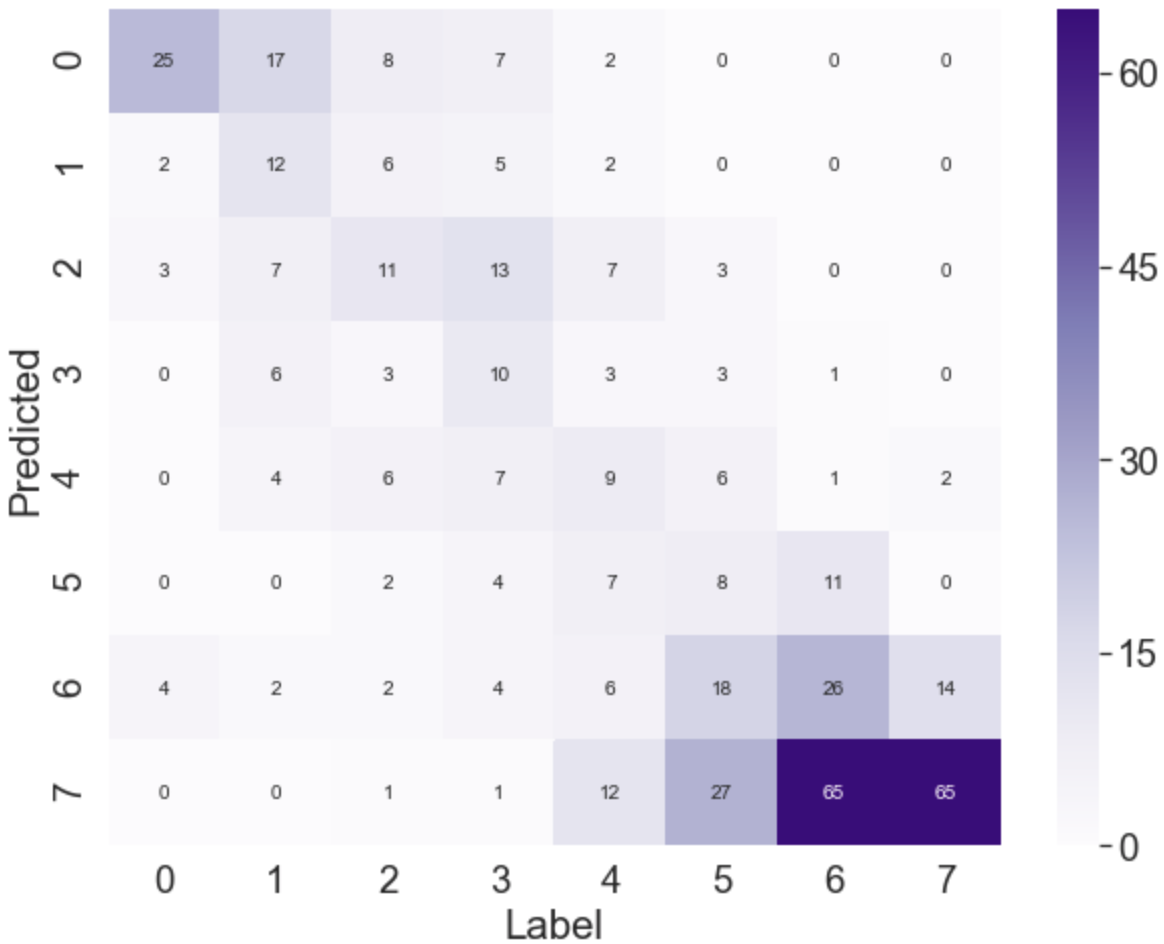}
\caption{Confusion matrix for the ruler scores of 0 ... 7 predicted by the single-branch noise model on the standard test set, with a ruler score error of 1.066.}
\label{mat8}
\end{figure}

Fig. \ref{mat8} is the confusion matrix for the same model on the eight ruler scores. Like the practical quality distribution, we have more samples on the good end in the test set, so counts are concentrated at the bottom right corner. The model is less accurate on images with median ruler scores. This issue is also observed when using other methods. It is partially due to the way we define the ruler score: the regions of scalars that fall into the median ruler scores are smaller than those for the end ones, so the error in ruler score caused by the same amount of error in raw scores can be larger. 

More experimental results of our model, including sample test images from each region in the two confusion matrices, and samples of saliency maps are included in the Appendix.

\subsection{Dual-task training}
We compare a dual-task trained model with two single-task trained ones on the task of noise assessment and motion detection, respectively, with the standard and OOD test sets. 

The dual-task model's architecture is shown in Fig. \ref{architect}, with both the noise and motion branches. The single task models share the same architecture except having only one of the two branches. The shared layers are optimized for the noise single-task model on the standard test set. 

The training sets for the three models share the same original slices, with corresponding noise and/or motion injected versions.
During the dual-task training, each training sample is labeled with either a noise score or a motion class, and is only used to update the corresponding branch and the shared layers. We simply use 0.5 as the decision threshold for inference of the motion branch.
We find the generation of simulated motion samples essential to the motion detection performance. When the same model is trained on original images only, after rebalancing the good and bad classes in training, only 78.0\% accuracy is achieved on the standard test set, 11.9\% less than that trained on our standard dataset where a motion injected version is included with each original slice. 

Table \ref{t3} lists the results from three (i.e. single-task noise, single-task motion, dual-task) models on the four (i.e. standard noise and motion, OOD noise and motion) datasets. First for the noise assessment task on the standard test set, the dual-task model's accuracy is 1.3\% lower than that of the single-task model, but its ruler score error is slightly better. On the OOD test set, the dual-task model is consistently better than the single-task model, with a 2.9\% higher accuracy and a 0.09 smaller error. Then for the motion detection task, the dual-task model is 2\% better in accuracy than the single-task one on both test sets. 

Since dual-task training consistently improves the performance on the OOD test set, we conclude that adding auxiliary tasks and using multi-task training improves the generalizability of our standard model. 
There are two possible sources of this improvement: training the shared layers to serve for multiple tasks serves as a form of regularization; the images in noise (or motion) set are not completely motion (or noise) free, and the effect from the irrelevant artifact is better separated by the dual-task model.
Our current dataset assumes the existence of single artifacts only. No sample with coexisting significant artifacts is intentionally included. The multi-task framework has more potential working with coexisting artifacts.

\begin{table} [!tbp]
\renewcommand\arraystretch{1.2}
  \caption{Comparison of two standard single branch models and a dual-task model on the binary classification accuracy and absolute ruler score error. The models are tested on two standard test sets for noise and motion artifacts, and two OOD test sets for noise and motion, respectively. Better results are in bold.}
  \vspace{1.7mm}
  \centering
  \begin{tabular}{|p{1.6cm}|p{2.27cm}||p{1.6cm}|p{1.5cm}|}
    \hline
    Metric & Dataset  &  Single-task & Dual-task \\
    \hline
    Accuracy  & Motion-standard & 87.80\% & \textbf{89.83}\%  \\
              & Motion-OOD & 85.00\% &  \textbf{87.50}\% \\
    \cline{2-4}
              & Noise-standard & \textbf{91.06}\% & 89.79\%  \\
              & Noise-OOD & 86.43\% & \textbf{89.29}\% \\
    \hline
    Score error & Noise-standard & 1.066 & \textbf{1.034} \\
              & Noise-OOD & 1.236 & \textbf{1.143} \\
    \hline
  \end{tabular}
\label{t3}
\end{table}

\section{Discussion} \label{S4}
We have proposed a NR-IQA model for noise and motion related quality assessment of MRI, with a focus on the framework for the perceptual noise assessment task. Our framework is not only reference-free in the test phase but also the training phase. It does not require a human provided subjective label score like MOS for training, either. 
This framework of calibrated scores for training and image rulers for inference is limited to applicable types of artifacts or aspects of image quality. The feature being assessed has to be suitable for simulation and controllable through a variable having a one-on-one relation with it. In our case, this independent variable is the noise variance.  
When this condition on the feature of interest is met, our framework can also be applied to assess image qualities for modalities other than MRI. For example, the noise quality of CT scans which depends on the radiation dose. 

There is no undersampling in our 2D FSE data so a sum-of-square reconstruction is used for generating the training and ruler images. Noise characteristics in the image domain are different for undersampled scans after compressed sensing \citep{cs}, deep learning based \citep{nature,2020ISPM,mine}, or other nonlinear reconstructions.
The proposed quality assessment framework still applies to these scans, while the training and ruler images have to be replaced by images from the corresponding undersampled data and reconstruction method. This is one of our future works.



Our model is currently trained on six anatomic regions and all major contrasts of FSE. We have tested extending the supported scan type without retraining the model on an unseen anatomy, and conclude that only adding a corresponding image ruler in inference is a viable option. However, it is still better to include the scan type of interest in training.
The current trained model's performance when extending to non-FSE scans is not yet tested. 

We plan to extend our model to abdominal images. These images are more susceptible to motion artifacts, so the motion detection branch can be more important. By including abdominal images in our dataset, we can have more real samples of motion or even coexisting artifacts, which can improve the training of the motion branch and support our plan of testing the multi-task model's ability to identify coexisting artifacts.

Additional branches can be added to the current model for the assessment of other artifacts, such as fat suppression failure and artifacts caused by B0 field inhomogeneity. The effect of these additional branches on the current dual-task performance can be explored. Our model has more applications beyond monitoring clinical scan quality. Since the CNN model is a differentiable function, it can be used to train a reconstruction or denoising neural networks as in adversarial training \citep{gan}. It can also be used to tune conventional compressed sensing reconstruction \citep{cs} of MRI, where the noise level in a reconstructed image has a one-on-one relation with the regularization parameter.

\section{Conclusion} \label{S5}
This paper presents a multi-task framework for artifact-specific quality assessment of MRI, with a focus on the context-dependent perceptual noise level assessment. It proposes the use of label calibration, a combination of model and human provided labels, as a more efficient way of label generation than the conventional one-to-one or many-to-one subjective labeling. Our label generation method also comes with a better defined labeling task than subjective score labeling. This paper then proposes the use of image rulers for a customizable and concrete way of interpreting scalar scores and defining quality levels. Experimental results show that classes defined by scan-specific image rulers also lead to better classification accuracies. Finally it is shown that a dual-task trained model has better generalizability and performance than two separate standard single-task models. 

The proposed work has several applications, but this work has focused on the goal of automated determination of whether images are acceptable for the intended diagnostic task, and if not, suggest or execute solutions. The proposed work is being deployed on our clinical scanners for this purpose.

\section*{Acknowledgments}
This work is supported by GE Healthcare, NIH R01EB009690, NIH R01EB026136.
We thank the following individuals for their assistance throughout our study:
Dr. Xucheng Zhu (GE Healthcare), Cedric Yue Sik Kin (Stanford University), Dr. Ali Syed (Stanford University), Dr. Naeim Bahrami (GE Healthcare), Dr. Marcus T. Alley (Stanford University), Dr. Jesse Sandberg (Stanford University).

\bibliographystyle{model2-names.bst}\biboptions{authoryear}
\bibliography{refs}


\end{document}


\begin{flushleft}
\Large \textbf{Appendix A.} Image rulers 
\end{flushleft}

\begin{figure}[h]
\centering
\includegraphics[width=0.99\linewidth]{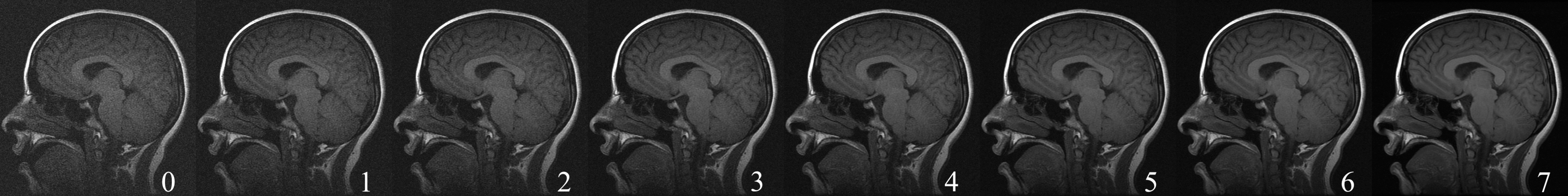}
\includegraphics[width=0.99\linewidth]{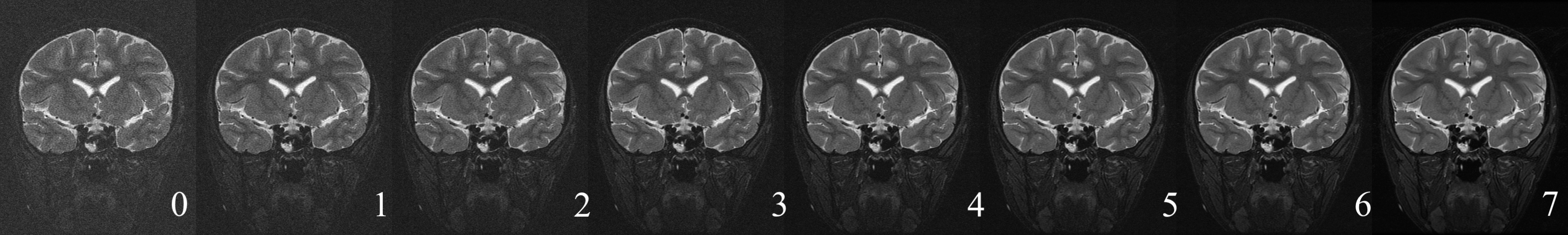}
\includegraphics[width=0.99\linewidth]{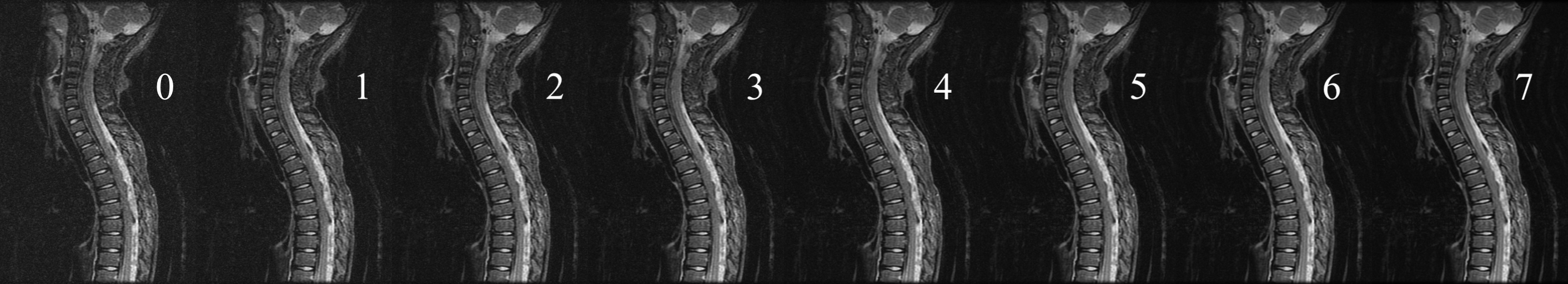}
\includegraphics[width=0.99\linewidth]{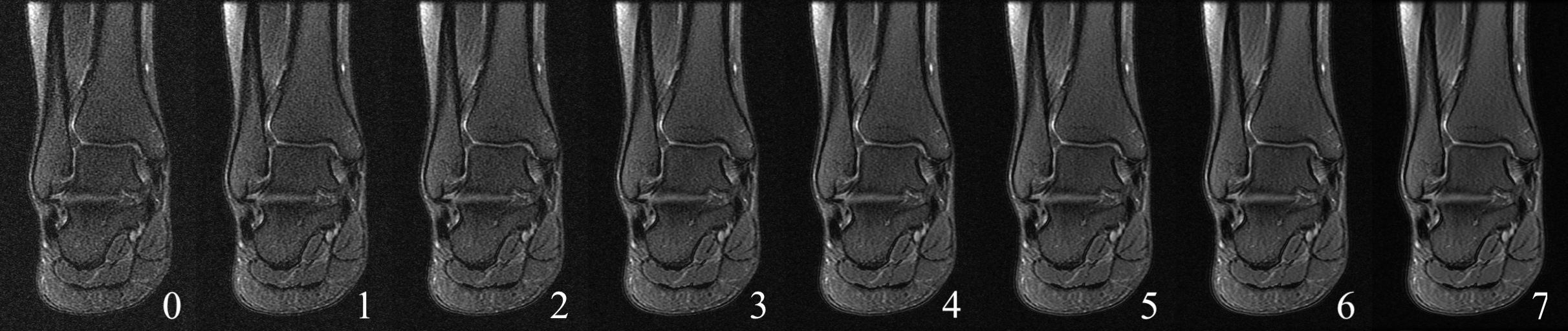}
\includegraphics[width=0.99\linewidth]{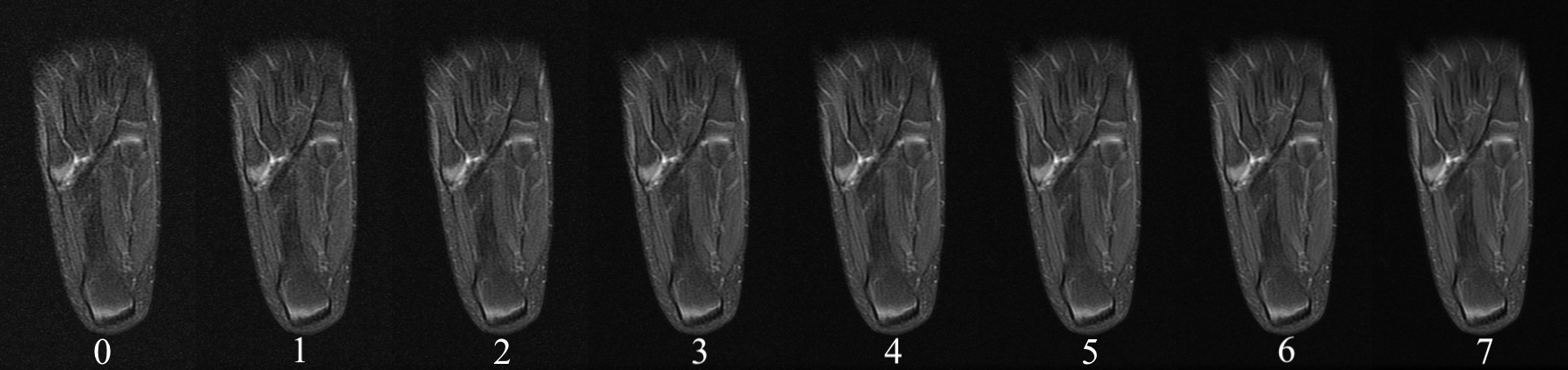}
\caption{\small\textbf{These five image rulers are, from top to bottom: NFS brain, FS brain, NFS spine, FS ankle, and FS foot. The thresholds chosen in these rulers from top to bottom are: 4.5, 3.5, 2, 3, and 3.} }
\label{r1}
\end{figure}

\begin{figure}[h]
\hspace*{-1cm}\includegraphics[width=1.1\linewidth]{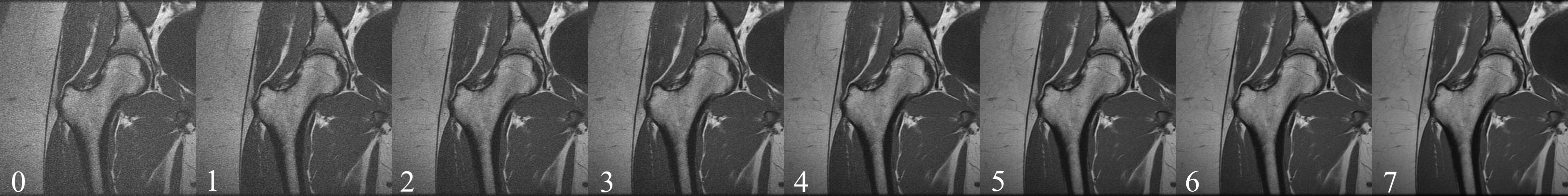}
\hspace*{-1cm}\includegraphics[width=1.1\linewidth]{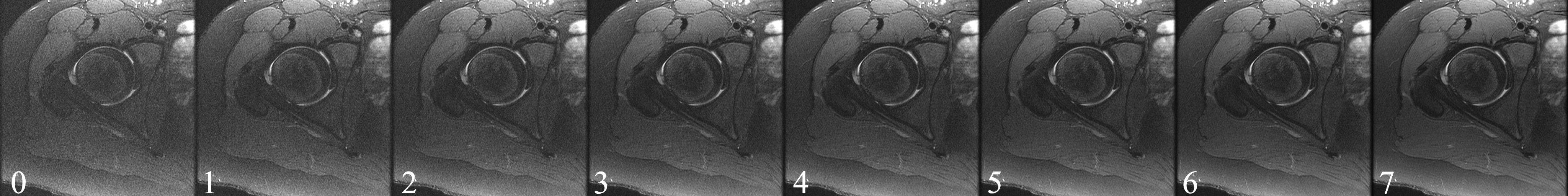}
\hspace*{-1cm}\includegraphics[width=1.1\linewidth]{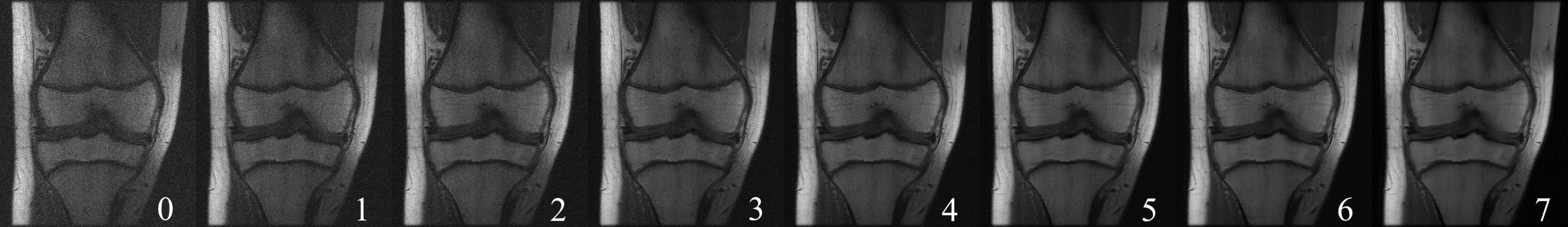}
\hspace*{-1cm}\includegraphics[width=1.1\linewidth]{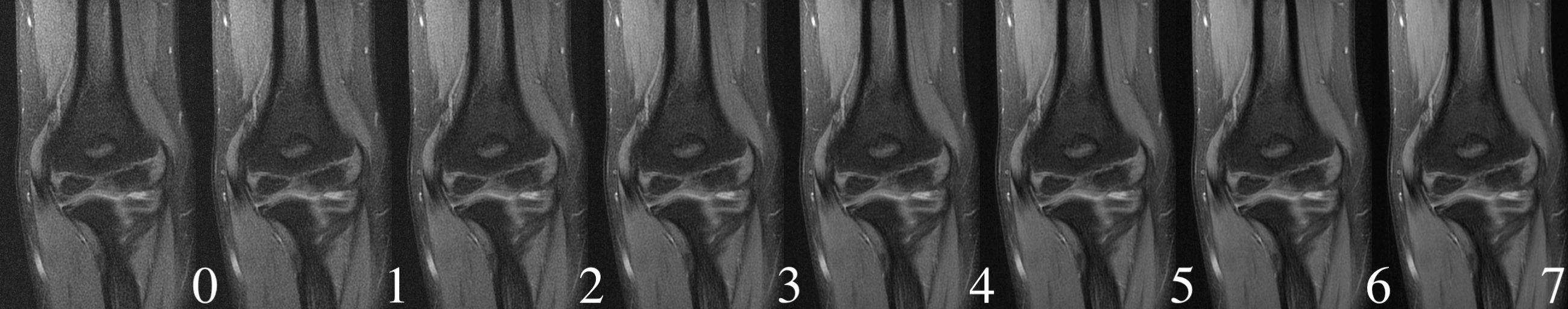}
\caption{\small\textbf{These four image rulers are, from top to bottom: NFS hip, FS hip, NFS knee, and FS elbow. The thresholds chosen in these rulers from top to bottom are: 3.5, 4.5, 5, and 4.5}}
\label{r2}
\vspace{.5cm}
\end{figure}

The nine image rulers and thresholds we use are shown in Fig. \ref{r1} and \ref{r2}. The rulers consist of seven anatomies. For spine, ankle and foot, only a FS or a NFS ruler is used because it is the only FS type for these anatomies in our default dataset. The FS and NFS rulers from elbow and knee, respectively, are shared among all elbow and knee scans because these two anatomies look similar. 
\vspace{3mm}

\clearpage

\begin{flushleft}
\Large \textbf{Appendix B.} DN vs. ReLU
\end{flushleft}

We compare three combinations of nonlinearity and normalization for the activation after convolutional layers. The divisive normalization (DN)  (Lyu and Simoncelli,2008), the ReLU with batch normalization (BN) (Ioffe and Szegedy, 2015) or layer normalization (LN) (Ba et al., 2016) are compared. Table \ref{at1} lists the results from single-tasks models within which the activation layers are the same. We find DN is best for the noise assessment task and ReLU with BN is best for the motion detection task. 

\begin{table} [!htbp]
\renewcommand\arraystretch{1.2}
  \caption{\small\textbf{Comparison of nonlinearities used for the activation layer after the convolutional layer. The models are tested on two default test sets for noise and motion artifacts. Best results are in bold.}}
  \vspace{.3cm}
  \centering
  \begin{tabular}{|c||c|c|c|}
    \hline
    Activation  & Motion Accuracy & Noise Accuracy & Noise score error \\
    \hline
    DN &     87.46\%   & \textbf{91.06}\%  & \textbf{1.066} \\
    ReLU+BN & \textbf{88.81}\% &  88.94\%  & 1.074 \\
    ReLU+LN &      87.12\%     &  89.36\%  & 1.085 \\
    \hline
  \end{tabular}
\label{at1}
\end{table}

\newpage

\begin{flushleft}
\Large \textbf{Appendix C.} Training details
\end{flushleft}

We use a mini-batch size of 10, and all five (for noise) or two (for motion) versions of a unique slice are contained in the same mini-batch. 
The training objectives are the root mean square error (between the CNN raw output and label score) and the binary cross-entropy loss for the noise and motion branches, respectively.
We use the Adam optimizer with a learning rate of $10^{-4}$, $\beta_1=0.9$, and $\beta_2=0.999$.   
We normalize each input image to zero mean and unit variance.
15\% of the inputs to the last fully-connected layer is dropped during (only) the training as a form of regularization. 

\begin{figure}[!h]
\includegraphics[width=0.32\linewidth]{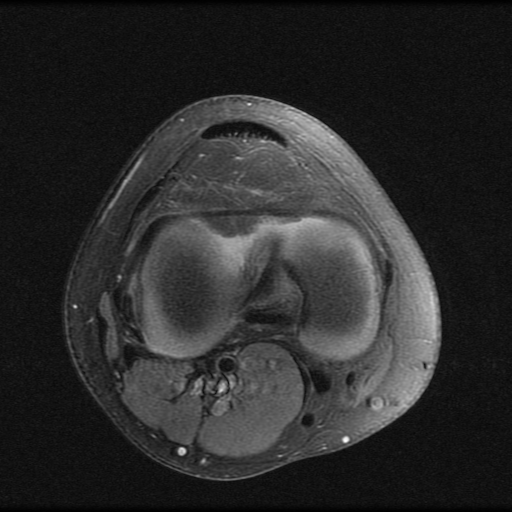}
\includegraphics[width=0.32\linewidth]{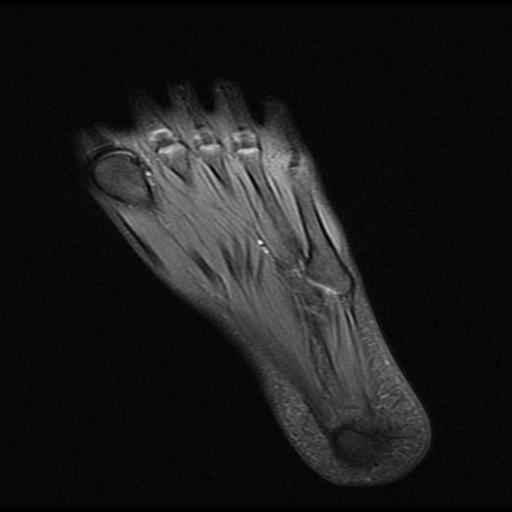}
\includegraphics[width=0.32\linewidth]{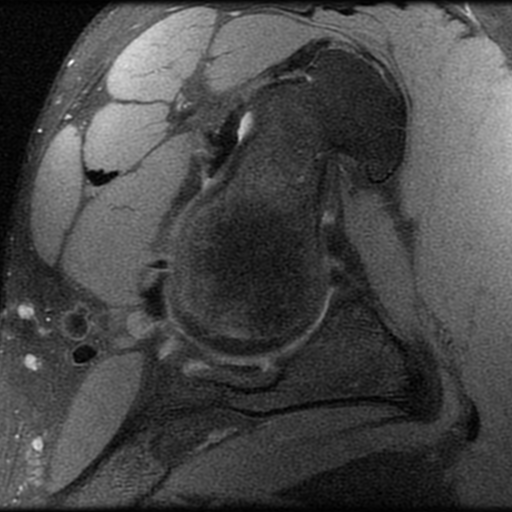}\\
\includegraphics[width=0.32\linewidth]{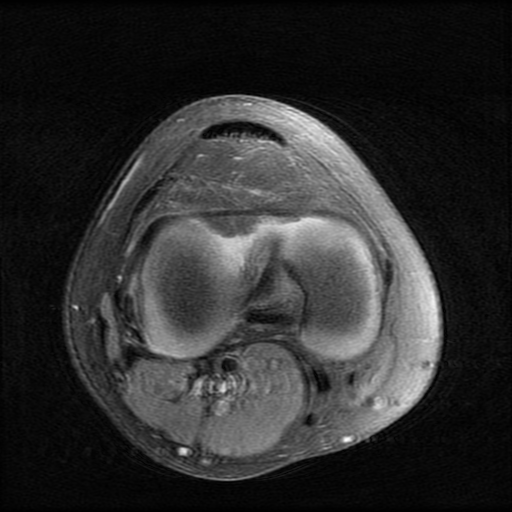}
\includegraphics[width=0.32\linewidth]{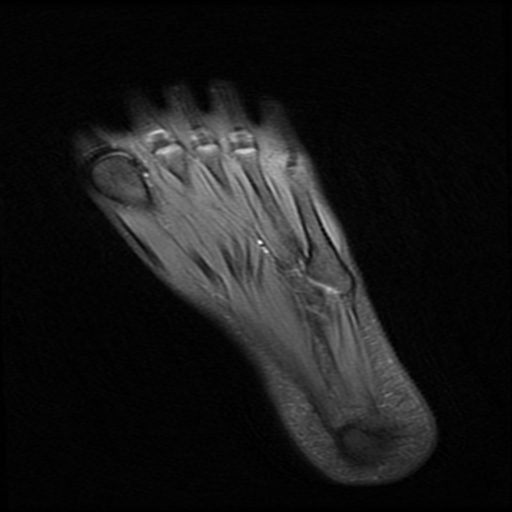}
\includegraphics[width=0.32\linewidth]{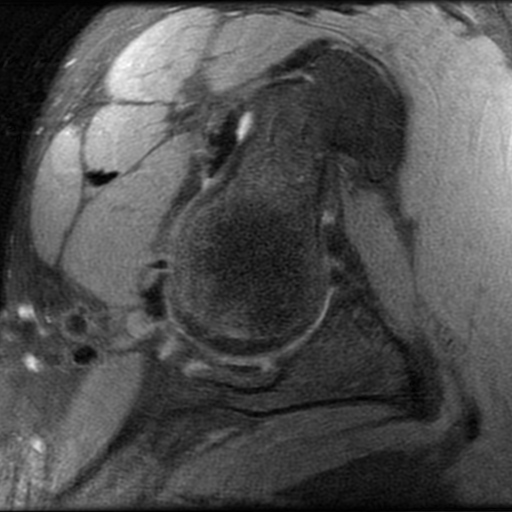}
\caption{\small\textbf{Samples of motion injected images. Images in the top row are the original versions, and images in the bottom row are the simulated versions. }}
\label{msample}
\end{figure}

\vspace{1.5mm}
We use motion simulated samples in the training. Fig. \ref{msample} presents three samples of our simulation outcomes. 

\vspace{1cm}

\newpage

\begin{flushleft}
\Large \textbf{Appendix D.} Test samples
\end{flushleft}

\begin{figure}[!h]
Prediction/label: \hspace{.1cm} 5, good \hspace{3.2cm} 7, good \hspace{4.3cm} 1, bad \\
\vspace{-3mm}
\includegraphics[width=0.32\linewidth]{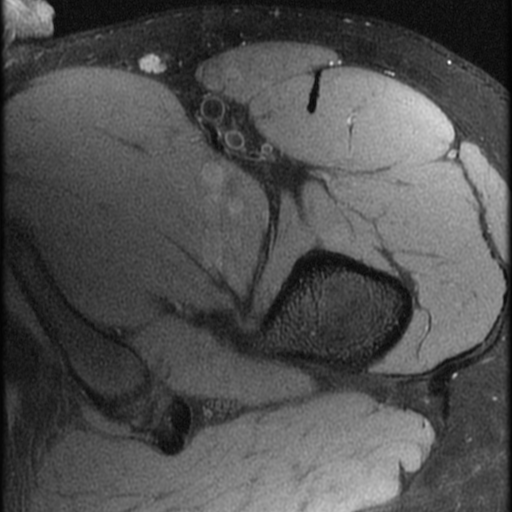}
\includegraphics[width=0.32\linewidth]{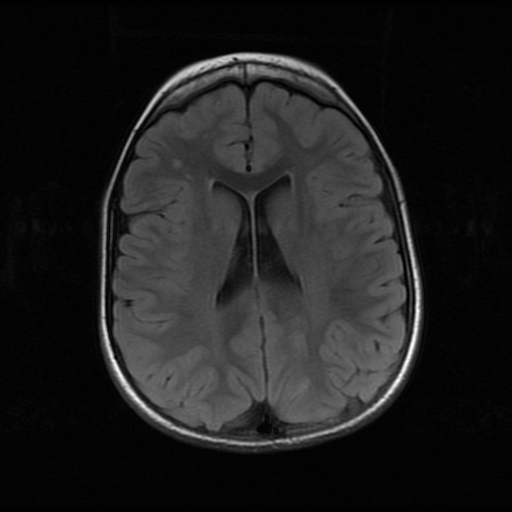}
\includegraphics[width=0.32\linewidth]{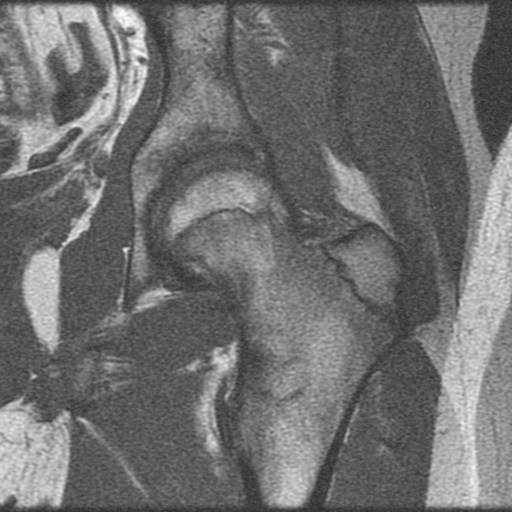}
\caption{\small\textbf{Correctly predicted test samples for the noise assessment task. }}
\label{c1}
\end{figure}

\begin{figure}[!h]
\begin{flushleft}
Prediction: \hspace{.4cm} 4, good \hspace{3.8cm} 5, good \hspace{4.2cm} 7, good \\
Label:      \hspace{1.2cm} 3, bad \hspace{4cm} 4, bad \hspace{4.4cm} 4, bad \\
\vspace{-3mm}
\end{flushleft}
\centering
\includegraphics[width=0.32\linewidth]{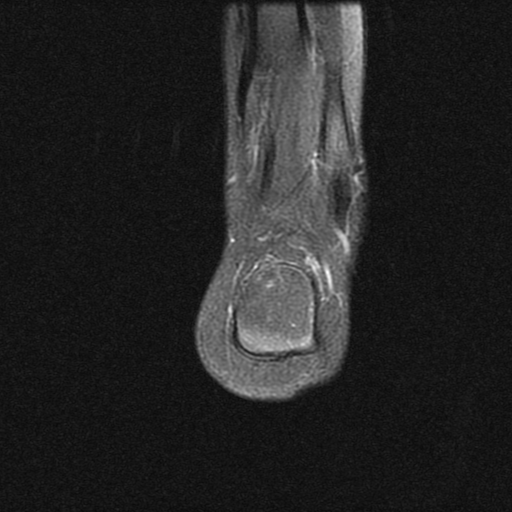}
\includegraphics[width=0.32\linewidth]{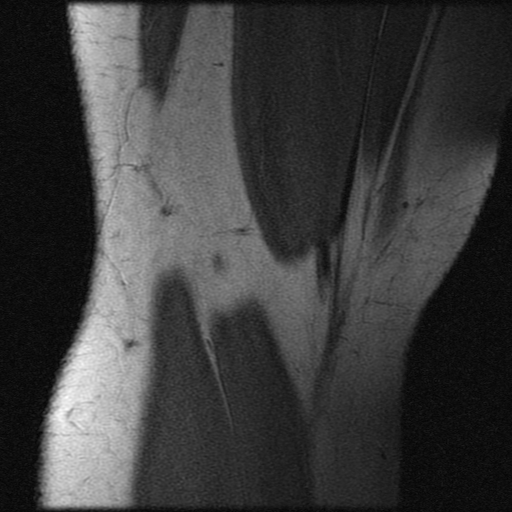}
\includegraphics[width=0.32\linewidth]{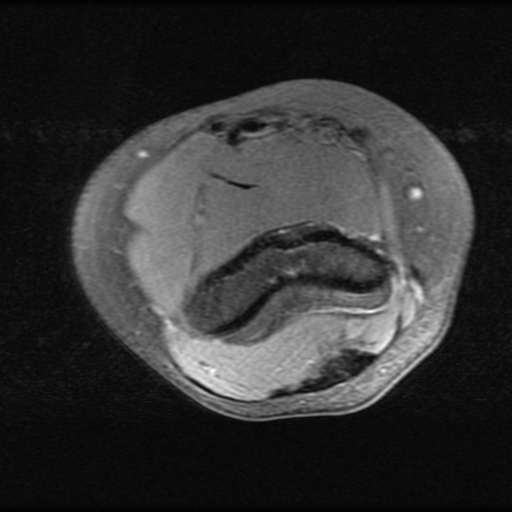}
\caption{\small\textbf{False positive test samples for the noise assessment task with their labeled and predicted rulers scores. }}
\label{c2}
\end{figure}

\begin{figure}[!h]
\begin{flushleft}
Prediction: \hspace{.5cm} 3, bad \hspace{4.2cm} 4, bad \hspace{4.4cm} 2, bad \\
Label:     \hspace{1.2cm} 4, good \hspace{4.cm} 5, good \hspace{4.2cm} 5, good \\
\vspace{-3mm}
\end{flushleft}
\centering
\includegraphics[width=0.32\linewidth]{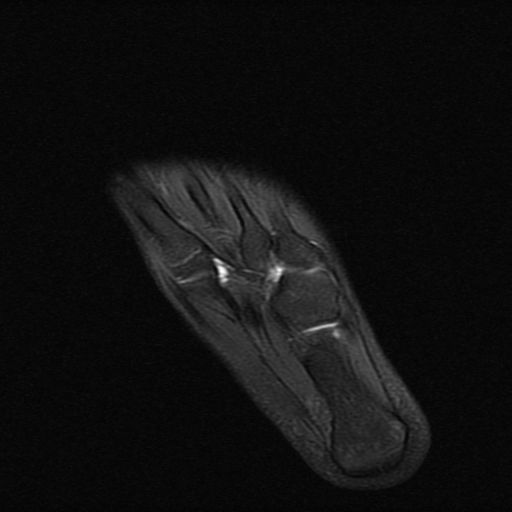}
\includegraphics[width=0.32\linewidth]{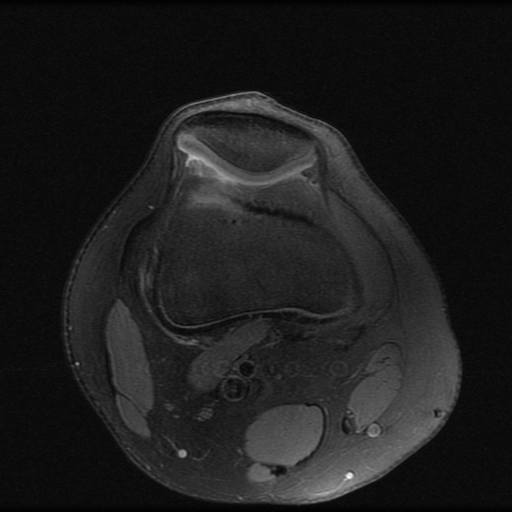}
\includegraphics[width=0.32\linewidth]{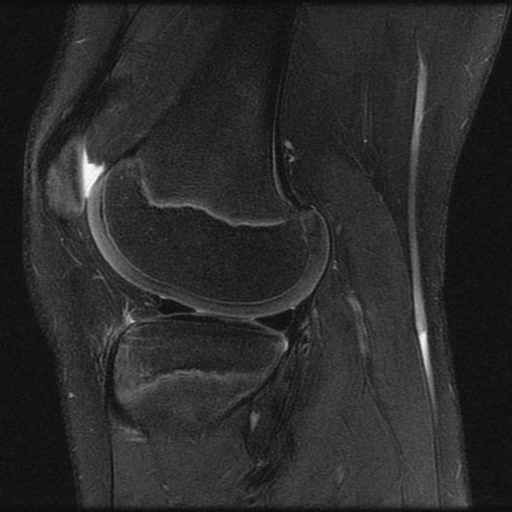}
\caption{\small\textbf{False negative test samples for the noise assessment task with their labeled and predicted rulers scores. }}
\label{c3}
\end{figure}

\begin{figure}[!h]
Prediction/label: \hspace{.2cm} good \hspace{3.7cm} bad \hspace{4.7cm} bad \\
\vspace{-3mm}
\includegraphics[width=0.32\linewidth]{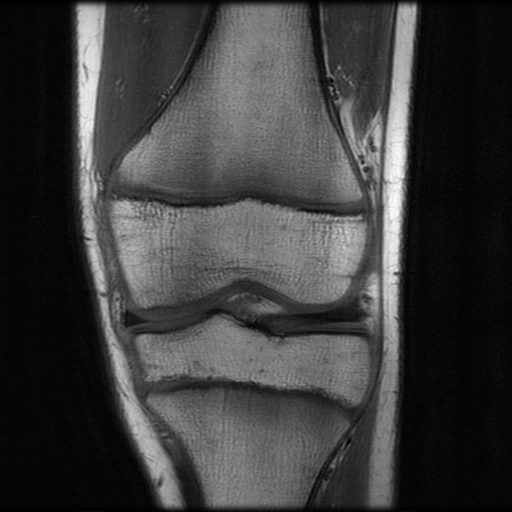}
\includegraphics[width=0.32\linewidth]{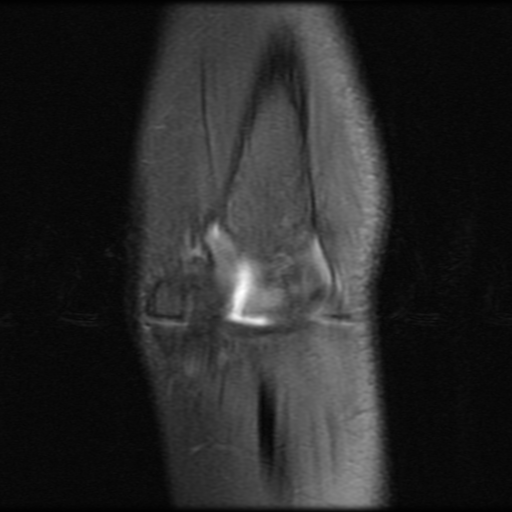}
\includegraphics[width=0.32\linewidth]{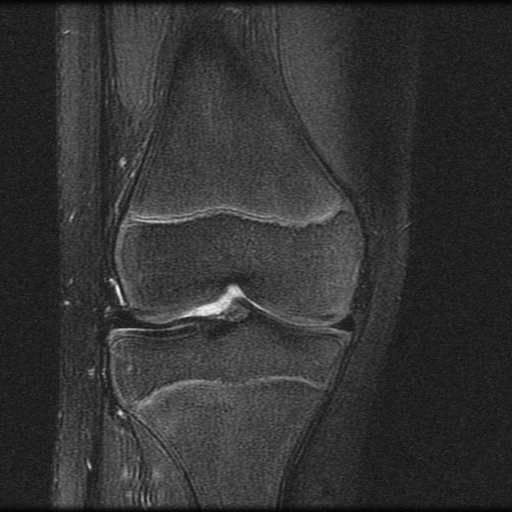}
\caption{\small\textbf{Correctly predicted test samples for the motion task. }}
\label{c4}
\end{figure}

\begin{figure}[!h]
\begin{flushleft}
Prediction: \hspace{.7cm}  good \hspace{4.2cm}   good \hspace{4.5cm}   good \\
Label:      \hspace{1.5cm}  bad \hspace{4.4cm}   bad \hspace{4.7cm}  bad \\
\vspace{-3mm}
\end{flushleft}
\centering
\includegraphics[width=0.32\linewidth]{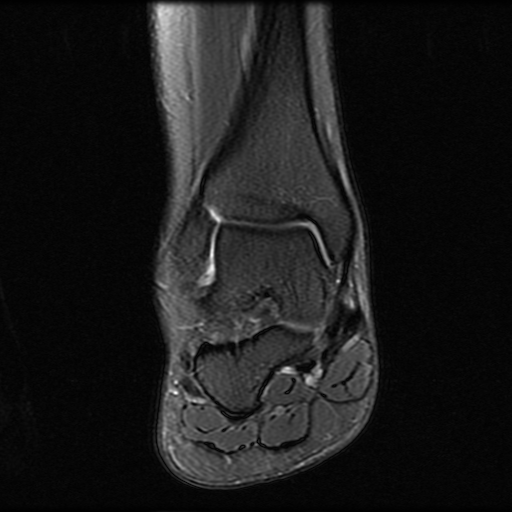}
\includegraphics[width=0.32\linewidth]{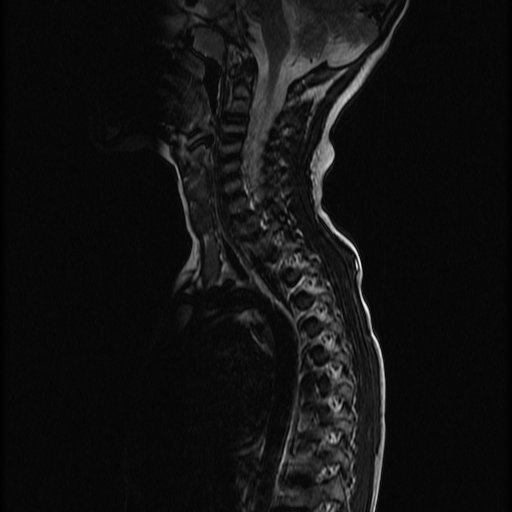}
\includegraphics[width=0.32\linewidth]{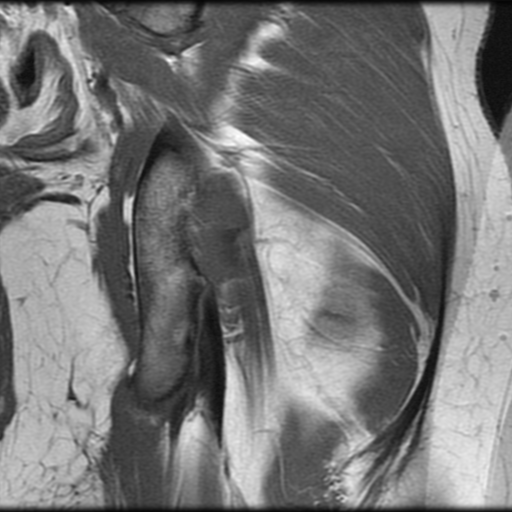}
\caption{\small\textbf{False positive test samples for the motion detection task. }}
\label{c5}
\end{figure}

\begin{figure}[!h]
\begin{flushleft}
Prediction: \hspace{.6cm} bad \hspace{4.6cm} bad \hspace{4.7cm} bad \\
Label:      \hspace{1.3cm}  good \hspace{4.4cm} good \hspace{4.5cm} good \\
\vspace{-3mm}
\end{flushleft}
\centering
\includegraphics[width=0.32\linewidth]{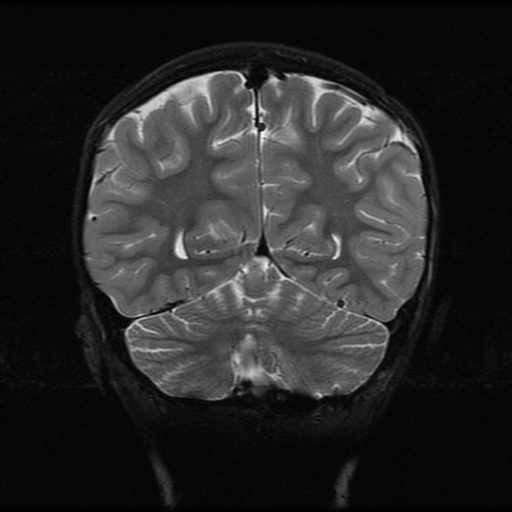}
\includegraphics[width=0.32\linewidth]{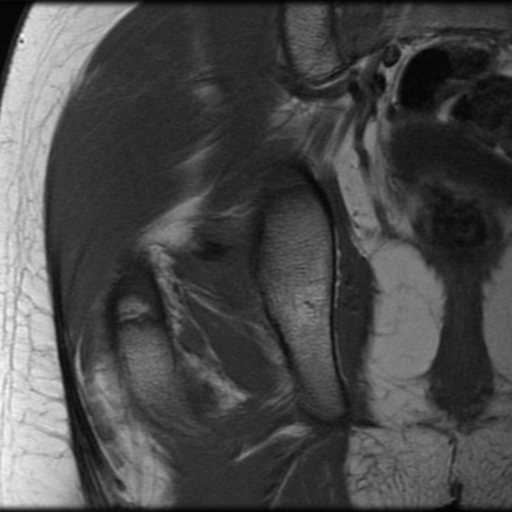}
\includegraphics[width=0.32\linewidth]{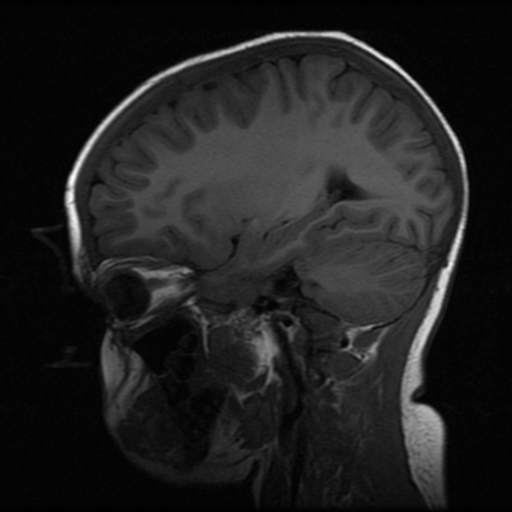}
\caption{\small\textbf{False negative test samples for the motion detection task. }}
\label{c6}
\end{figure}

We present samples of test images in the default dataset with their labels and predictions from the dual-task model. Fig. \ref{c1}, \ref{c2}, and \ref{c3} present three true, false positive, and false negative samples, respectively for the noise assessment task. 
Fig. \ref{c4}, \ref{c5}, and \ref{c6} present three true, false positive, and false negative samples, respectively for the motion detection task.
\vspace{3mm}

\clearpage

\clearpage

\clearpage

\begin{flushleft}
\Large \textbf{Appendix E.} Saliency maps
\end{flushleft}

\begin{figure}[!h]
\begin{flushleft}
\hspace{2.2cm} image \hspace{3.9cm} saliency map \hspace{3.7cm} overlaid \\
\vspace{-3mm}
\end{flushleft}
\centering
\includegraphics[width=0.32\linewidth]{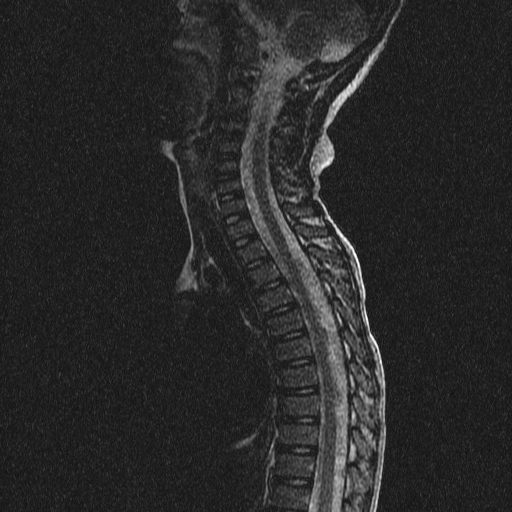}
\includegraphics[width=0.32\linewidth]{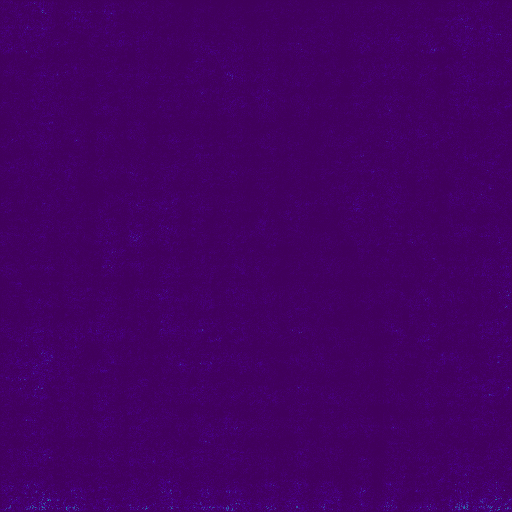}
\includegraphics[width=0.32\linewidth]{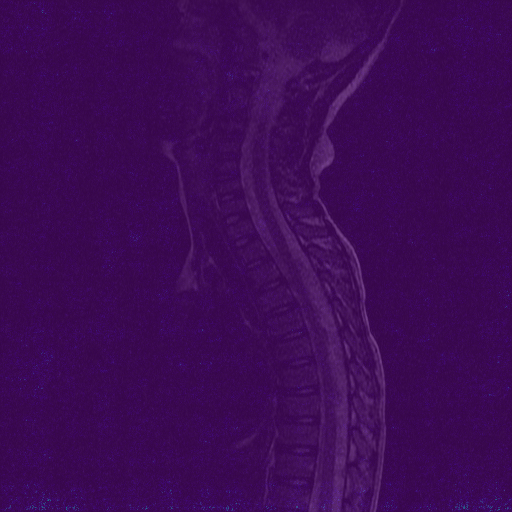}
\vspace{-.3cm}
\caption{\small\textbf{A test samples and its saliency map with respect to the noise score.}}
\vspace{-.3cm}
\label{e1}
\end{figure}

\begin{figure}[!h]
\begin{flushleft}
\hspace{2.2cm} image \hspace{3.9cm} saliency map \hspace{3.7cm} overlaid \\
\vspace{-3mm}
\end{flushleft}
\centering
\includegraphics[width=0.32\linewidth]{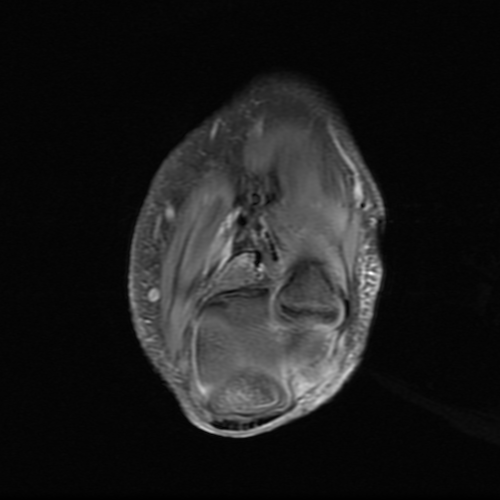}
\includegraphics[width=0.32\linewidth]{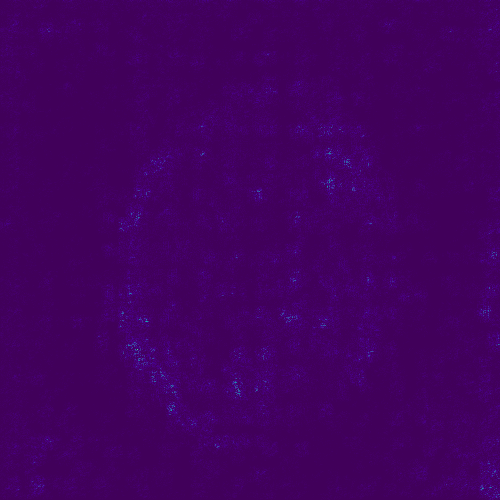}
\includegraphics[width=0.32\linewidth]{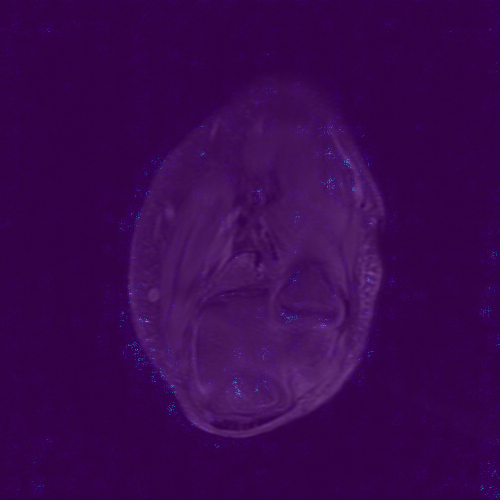}\\
\vspace{.2cm}
\includegraphics[width=0.32\linewidth]{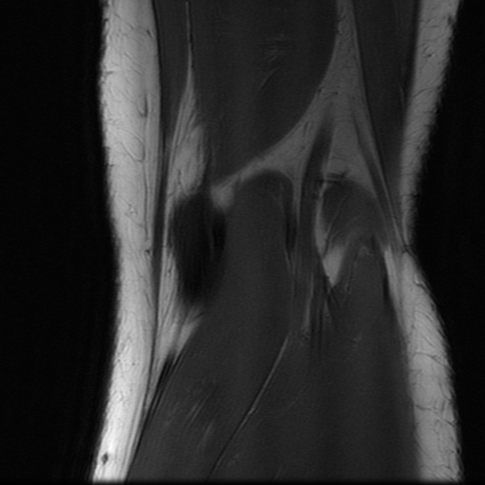}
\includegraphics[width=0.32\linewidth]{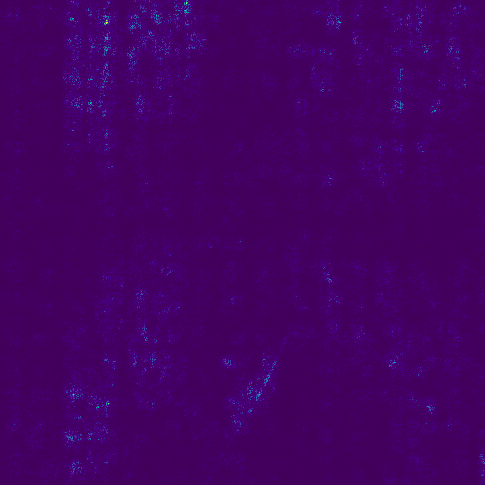}
\includegraphics[width=0.32\linewidth]{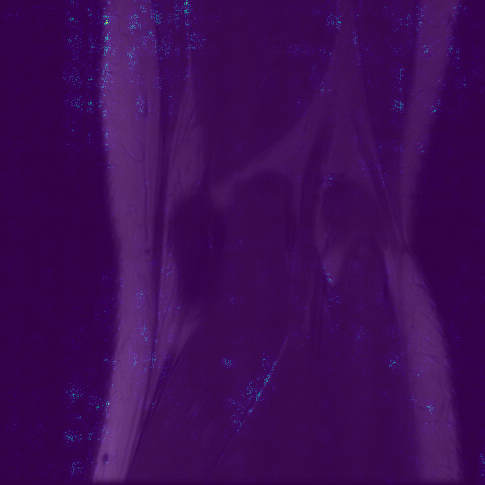}
\vspace{-.3cm}
\caption{\small\textbf{Two test samples and their saliency maps with respect to the motion probability.}}
\label{e2}
\end{figure}

Fig. \ref{e1} and \ref{e2} present test images with their saliency maps from the dual-task model. Patterns of the saliency maps are consistent with the distribution of the artifacts. Noise spreads out relatively uniformly over the whole image. Motion exists within and near the edge of the object.

